\documentclass[preprint,12pt]{elsarticle}

\usepackage{graphicx}
\usepackage{subcaption}
\usepackage{amssymb}
\usepackage{amsmath}
\usepackage{comment}
\usepackage{color}
\usepackage{soul}
\usepackage{booktabs}
\usepackage{tabularx}
\usepackage{siunitx}
\usepackage{algorithm}
\usepackage{algcompatible}
\usepackage[left=1in, right=1in]{geometry}
\usepackage{url}
\usepackage{hyperref}
\usepackage{setspace}
\usepackage{lineno}
\begin{document}
%\linenumbers
\begin{frontmatter}

\title{A numerical study on the effect of rolling friction on clogging of pores in particle-laden flows}

\author[1,2,3]{Sagar G. Nayak} 
\author[4]{Zhenjiang You} 
\author[2]{Yuchen Dai} 
\author[2]{Geoff Wang} 
\author[3]{Prapanch Nair\corref{cor1}}

\affiliation[1]{organization={University of Queensland - Indian Institute of Technology Delhi (UQ-IITD) Research Academy},%Department and Organization
            addressline={Hauz Khas}, 
            city={New Delhi},
            postcode={110016}, 
            country={India}}
\affiliation[2]{organization={School of Chemical Engineering, The University of Queensland},%Department and Organization
            addressline={St. Lucia}, 
            city={QLD},
            postcode={4072}, 
            country={Australia}}
\affiliation[3]{organization={Department of Applied Mechanics, Indian Institute of Technology Delhi},%Department and Organization
            addressline={Hauz Khas}, 
            city={New Delhi},
            postcode={110016}, 
            country={India}}
\affiliation[4]{organization={School of Petroleum, China University of Petroleum - Beijing at Karamay},%Department and Organization
            addressline={Karamay}, 
            city={Xinjiang},
            postcode={834000}, 
            country={China}}
\cortext[cor1]{Corresponding author: pnair@am.iitd.ac.in; Tel: +91 11 2659 1226}

\begin{abstract}
 Particulate matter in a fluid injected into a porous reservoir  impairs its permeability spatio-temporally due to pore clogging. As particle volume fraction increases near the pore throats, inter-particle contact mechanics determine their jamming and subsequent pore clogging behavior. During contact of particles submerged in a fluid, in addition to sliding friction, a rolling resistance develops due to a several micromechanical and hydrodynamic factors. A coefficient of rolling friction is often used as a lumped parameter to characterize particle rigidity, particle shape, lubrication and fluid mediated resistance, however its direct influence on the clogging behavior is not well studied in literature. We study the effect of rolling resistance on the clogging behavior of a dense suspension at pore scale using direct numerical simulations (DNS). A discrete element method (DEM) library is developed and coupled with an open-source immersed boundary method (IBM) based solver to perform pore and particle resolved simulations. Several 3D validations are presented  for the DEM library and the DEM-IBM coupling and the effect of rolling resistance on clogging at a pore entry is studied.
\end{abstract}


\end{frontmatter}

%% main text
\section{Introduction} \label{sec:intro}

Particle-laden flows through constricted pathways are common in natural and industrial processes. A key concern in these systems is the retention and accumulation of particles at unfavorable locations, commonly referred to as clogging.Clogging causes irreversible decline in permeability (also known as formation damage \cite{civan2023reservoir}) in subsurface applications like geothermal reservoirs \cite{you_fines_2019}, oil and natural gas extraction fields \cite{krueger_overview_1986} and carbon storage sites, significantly affecting the fluid transport. Formation damage increases the pumping power required to transport fluid \cite{yuan_comprehensive_2018}, leading to increased operational cost and possible abandonment of the reservoir. An understanding of the clogging mechanism is crucial for efficient and reliable design of control mechanisms.

A particle migrating through a pore channel is brought to rest under one or more of the three circumstances: (i) if the singular particle is larger than the channel width in which case it is `strained' \cite{herzig_flow_1970}, (ii) if multiple small particles arrive simultaneously forming a `bridge' across the channel \cite{ramachandran_plugging_1999}, and 
(iii) particles attach to the pore walls due to electrostatic attraction and  result in progressive reduction of the channel width\cite{katz1987prediction}. 
The second mechanism among these, namely bridging, causes abrupt flow obstruction, nonlinearly  governed by the flow rate, particle volume fraction and particle to constriction size ratio \cite{he_pore-scale_2025, mondal_coupled_2016}, and is difficult to predict.

Most experimental investigations are limited to measurement of inlet to outlet particle concentration changes, since visualization of clogging is difficult due to three dimensionality and the opaqueness of the system.  
While these measurements can estimate the constrictions inside the system \cite{chalk_pore_2012}, they do not reveal the actual clogging mechanism. 
Two-dimensional transparent microfluidic cells are used to examine clogging of particles in bottlenecks \cite{sendekie2016colloidal}, nozzles \cite{vani_role_2024} and analogous porous structures \cite{he_pore-scale_2025}, recently. However, their conclusions can be incomplete due to the absence of the third dimension. Mathematical models are also widely employed for different applications \cite{yang_stochastic_2020, you_mathematical_2016, jegatheesan2005deep}. However, they lack the particle scale description of clogging mechanism. 

At particle and pore scale, various numerical methods like fully Eulerian \cite{li_coupled_2020}, fully Lagrangian \cite{nelson2011new}, and hybrid methods \cite{tsuji_discrete_1993} have been used to study particulate transport. Among the hybrid methods, the two-way coupling of discrete element method (DEM)\cite{cundall1979discrete} with traditional computational fluid dynamics (CFD) solvers has been a popular choice \cite{goniva2012influence}. In this CFD-DEM method, the hydrodynamic forces  decide the particle trajectories, while the particles disturb the local flow field. In CFD-DEM, the particle shape may either be well resolved by the CFD mesh or be modeled as subgrid sized point particles. 
While unresolved solvers use a momentum exchange term to communicate between particle and fluid, resolved solvers integrate fluid forces over the particle surface to compute the hydrodynamic forces. 
Even with unresolved CFD-DEM simulations,
Shahzad \textit{et al.} \cite{shahzad2018aggregation} concluded that larger particle aggregates lead to clogging in 3D compared to 2D. Unresolved methods have also been used to capture clogging and particle retention in rocks \cite{mirabolghasemi_prediction_2015}, packed beds \cite{wang2023unresolved}, constricted channels \cite{mondal_coupled_2016} and membrane fouling process \cite{lohaus_what_2018}. Although inexpensive compared to resolved techniques, unresolved methods lead to flawed flow field predictions \cite{liu_mechanisms_2024}. 

Particle-resolved CFD-DEM is also widely used. Kermani \textit{et al.} \cite{samari_kermani_direct_2020}  studied particle transport in a 2D constricted channel using particle resolved simulations. Although simulations shed light on the events that lead to clogging, they fall short in predicting post-clog flow fields, since a 2D clog blocks the flow completely compared to a 3D blocked scenario which permits interstitial flow. There have been recent 3D studies using particle resolved simulations on clogging in constricted channels \cite{mondal_coupled_2016}, representative pore structure \cite{liu_mechanisms_2024} and porous media \cite{elrahmani_clogging_2023}. In these studies parameters like flow rate, particle concentration, particle to throat size ratio, ionic strength etc. were accounted for.  

At small length scales ($<\SI{100}{\mu m}$), non-contact forces like van der Waal forces and electrostatic forces become prominent \cite{zheng_interparticle_2024} and trigger particle bridging and eventual clogging \cite{di_vaira_hydrodynamic_2023, samari_kermani_direct_2020, liu_mechanisms_2024}. However, at larger length scales ($>\SI{100}{\mu m}$)  these forces are negligible and the mechanism responsible for bridging and clogging remains underexplored. 

Studies on dry granular flows through hopper-like systems show that rolling resistance between particles and between particles and walls plays a dominant role in jamming near the throat \cite{sakaguchi1993plugging, balevicius_effect_2012}. In dense suspensions, while some studies have considered the effect of sliding friction \cite{li_coupled_2020}, the influence of rolling on clogging remains unexplored. In this paper, we address this gap by studying the role of rolling resistance on pore clogging. 

We begin by discussing the development  of an immersed boundary method (IBM)-DEM methodology in \S \ref{sec:methodology}.
IBM helps in handling fully resolved flows around particle geometries and through pores and DEM helps in modeling particle-particle and particle-wall interactions. The IBM methodology is implemented using the open-source  Immersed Boundary Adaptive Mesh Refinement library (IBAMR) \cite{griffith_adaptive_2018}. To this, we integrate an in-house developed DEM library to realize the proposed IBM-DEM methodology. This library is also made available as open source for the benefit of the reader.
Then, some canonical tests are performed in \S \ref{sec:validAndGridInd} to show the reliability of the formulation. In \S \ref{sec:results} test cases are formulated for studying the effect of rolling friction on pore clogging and the results are  discussed. In \S \ref{sec:conclusion} we conclude the paper by listing the insights gained from the study.

\section{Methodology}\label{sec:methodology} 

In this section, we describe the mathematical formulation of the proposed IBM-DEM technique. We begin with the governing equations for the coupled fluid-structure interaction and their numerical solution using the immersed boundary method. Different types of contact forces involved in particulate flows and their computation are then explained. We conclude the section with a discussion on coupling of the DEM solution with the fluid flow governing equations. We use bold letters and symbols for vector quantities and regular letters and symbols for scalars.

\subsection{Immersed Boundary Method} \label{sec:cfd-ibm}

The governing equations for the coupled fluid-structure interaction for an isothermal incompressible fluid interacting with rigid solids, in the Eulerian frame of reference, are 
\begin{equation}\label{eqn:momentum}
    \rho_f \left(\frac{\partial \mathbf{u}(\mathbf{x},t)}{\partial t} + (\mathbf{u}(\mathbf{x},t) \cdot \nabla)\mathbf{u}(\mathbf{x},t)\right) = -\nabla p(\mathbf{x},t) + \mu_f \nabla\cdot\nabla \mathbf{u}+ \mathbf{f}_b(\mathbf{x},t),
\end{equation}
and 
\begin{equation} \label{eqn:continuity}
    \nabla \cdot \mathbf{u}(\mathbf{x},t) = 0,
\end{equation}

in a fluid domain, $\mathcal{U}$ ($\mathcal{U} \in \mathbb{R}^3$), with appropriate flow boundary conditions at its  boundary $\partial \mathcal{U}$. $\mathcal{U}$ is divided into time-dependent subdomains namely $\mathcal{U}_f$ for fluid and $\mathcal{U}_b$ for solid, such that $\mathcal{U} = \mathcal{U}_f \cup \mathcal{U}_b$. Further, the Lagrangian description of the immersed structure is realized by the material coordinate system $\Omega_b$ ($\Omega_b \in \mathbb{R}^3$). In the equations \ref{eqn:momentum} and \ref{eqn:continuity}, $\mathbf{u}(\mathbf{x},t)$ and $p(\mathbf{x},t)$ represent velocity vector and pressure, respectively, at a position $\mathbf{x}$ at time $t$. $\rho_f$ is the uniform fluid density and $\mu_f$ is the uniform fluid viscosity. The body force per unit volume (force density) $\mathbf{f}_b(\mathbf{x},t)$ due to an immersed structure can be computed as

\begin{equation}
    \mathbf{f}_b(\mathbf{x},t) \equiv \sum_l  \mathbf{F}_b(\mathbf{s}_l,t) W(\mathbf{x} - \mathbf{X}(\mathbf{s}_l,t)) d\mathbf{s}_l \approx \int_{\Omega_b} \mathbf{F}_b(\mathbf{s},t) W(\mathbf{x} - \mathbf{X}(\mathbf{s},t)) d\mathbf{s}.
    \label{eqn:Eulerian_body_force}
\end{equation}

Here the structure is discretized into a set of marker points  $\{ \mathbf{s}_l\}_{l=1}^n$, where $l$ denotes the index of the marker points in $\Omega_b$. $\mathbf{F}_b(\mathbf{s}_l,t)$ is the force density specified at a discrete marker point $l$ located at $\mathbf{s}_l$ in the reference frame of the structure.  

Also, $W$ is a smoothed and normalized kernel function that approximates the Dirac delta function.
In this study, we state the body force $\mathbf{F}_b(\mathbf{s}_l,t)$ in the equation \ref{eqn:Eulerian_body_force} to be sum of two distinct forces similar to \cite{sharma_coupled_2022}. That is,
\begin{equation}
    \mathbf{F}_b(\mathbf{s}_l,t) = \mathbf{F}_c(\mathbf{s}_l,t) + \mathbf{F}_{\text{col}}(\mathbf{s}_l,t),
    \label{eqn:fb=fc+fcol}
\end{equation}
where, $\mathbf{F}_c(\mathbf{s}_l,t)$ is the force required to enforce the rigidity constraint in equation \ref{eqn:rigidity_constraint}. This force is computed using the Lagrange multipliers method \cite{shirgaonkar_new_2009, glowinski_distributed_1999}. We include $\mathbf{F}_{\text{col}}$ to account for the forces generated due to particle-particle and particle-wall collisions. 

Velocity of the Lagrangian markers can be obtained from the Eulerian reference frame of the fluid as,
\begin{equation}
    \mathbf{U}(\mathbf{s}, t)=\int_{\mathcal{U}_{\mathrm{b}}} \mathbf{u}(\mathbf{x}, t) \delta(\mathbf{x}-\mathbf{X}(\mathbf{s}, t)) \mathrm{d} \mathbf{x}.
    \label{eqn:Lagrangian_velocity}
\end{equation}

Here, the volume $\mathcal{U}_b$ is assumed to be filled with a fictitious fluid that obeys the constraint of rigid body motion \cite{shirgaonkar_new_2009}. That is,
\begin{equation}
    \frac{1}{2}\left[\nabla \mathbf{u}+\nabla \mathbf{u}^T\right]=\frac{1}{2}\left[\nabla \mathbf{u}_{\mathrm{k}}+\nabla \mathbf{u}_{\mathrm{k}}^T\right] \quad \text { in } \mathcal{U}_{\mathrm{b}},
    \label{eqn:rigidity_constraint}
\end{equation}
where, $\mathbf{u}_{\mathrm{k}} = \mathbf{u}_{\mathrm{k}}(\mathbf{s},t)$ is the prescribed velocity of the body. For a neutrally buoyant body whose motions are unconstrained, $\mathbf{u}_{\mathrm{k}}$ is the velocity resulting from the fluid drag. We assume a no-slip boundary condition unless otherwise stated. 

As mentioned previously, the immersed boundary method explained in this section is realized using the IBAMR library \cite{griffith_adaptive_2018}. For more information like discretization and software implementation, readers are referred to \cite{bhalla_unified_2013}. Further, using DEM \cite{poschel_computational_2005} we compute $\mathbf{F}_{\text{col}}$ in equation \ref{eqn:fb=fc+fcol}. The details of force computation using DEM and its integration with IBM are explained in the following sections.

\subsection{Discrete Element Method} \label{sec:dem}

\begin{figure}[htb]
    \centering
    \includegraphics[width=0.8\linewidth]{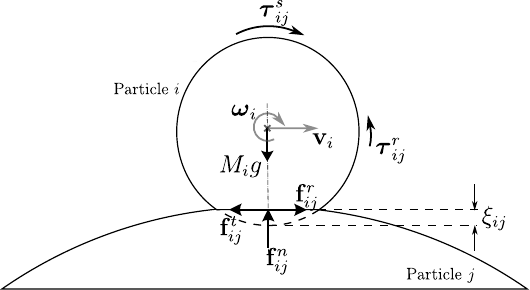}
    \caption{Forces and torques experienced by a particle during collision.}
    \label{fig:meth_collision}
\end{figure}

\begin{figure}[htb]
     \centering
     \includegraphics[width=0.75\linewidth]{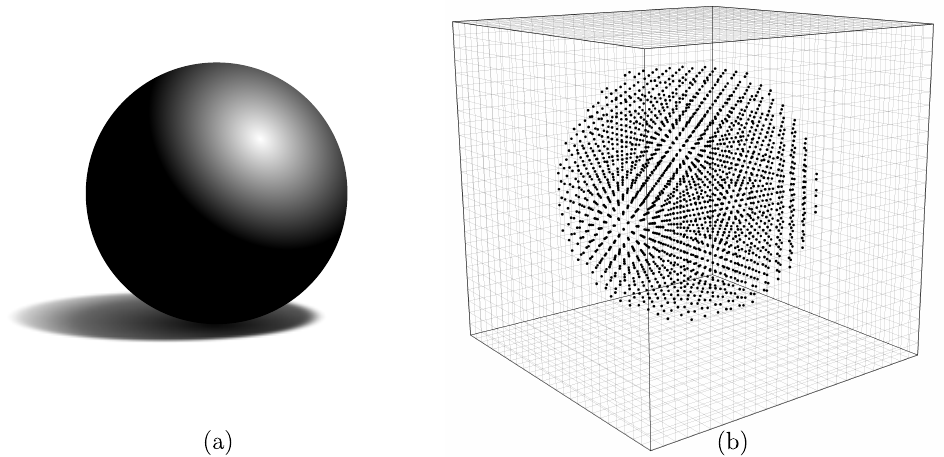}
     \caption{Interpretation of particle by (a)DEM and (b)IBM}
     \label{fig:particle in DEM and IBM}
 \end{figure}

\begin{algorithm}[htb!]
\caption{IBM-DEM flow of commands}
\begin{algorithmic}[1]
    \STATE $t \gets 0$
    \STATE Initialize $\mathbf{r}$, $\mathbf{u}(\mathbf{x})$, and $p(\mathbf{x})$.
    \STATE Fill particles and structures with Lagrangian markers.
    \WHILE{$t \leq t_{end}$,}
        \STATE Interpolate $\mathbf{u}$ onto markers, get $\mathbf{U}(\mathbf{s},t)$ (eq. \ref{eqn:Lagrangian_velocity}).
        \STATE Compute particle velocities ($\mathbf{v}$ and $\boldsymbol{\omega}$).
        %\STATE 
        \STATE $c = 0$
        \WHILE{$c \leq \text{ceil}\left(\frac{\Delta t_{\text{CFD}}}{\Delta t_{\text{DEM}}}\right)$}
            \STATE Initialize $\mathbf{f}_i\gets 0$, $\boldsymbol{\tau}_i\gets0$
            % Gear's algorithm
            \STATE Predict $\mathbf{r^*}$, $\mathbf{v^*}$, and $\boldsymbol{\omega^*}$
            \FOR{each particle $i$,}
                \FOR{each neighboring particle $j$}
                    \IF{ $\xi_{ij} > 0$}
                        \STATE Compute $\mathbf{f}_i$ and $\boldsymbol{\tau}_i$ using eq. \ref{eqn:Fn}-\ref{eqn:total force and torque}.
                        \rlap{\smash{$\left.\begin{array}{@{}c@{}}\\{}\\{}\\{}\\{}\\{}\\{}\\{}\\{}\\{}\\{}\\{}\\{}\end{array}\color{black}\right\}%
          \color{black}\begin{tabular}{l}DEM Loop\end{tabular}$}}
                    \ENDIF
                \ENDFOR
            \ENDFOR
            \STATE Correct $\mathbf{r}$, $\mathbf{v}$, and $\boldsymbol{\omega}$
            \STATE $c \gets c+1$
        \ENDWHILE
        \STATE
        \STATE Update markers' position ($\mathbf{s}$)
        \STATE Distribute force onto markers using eq. \ref{eqn:Ffil} - \ref{eqn:fcol=ffil+ftil}.
        \STATE Spread force from markers to Eulerian grid using eq. \ref{eqn:Eulerian_body_force}.
        \STATE Update $\mathbf{u}(\mathbf{x},t)$ and $p(\mathbf{x},t)$
        \STATE $t \gets t + \Delta t_{\text{CFD}}$
    \ENDWHILE
\end{algorithmic}
\label{alg:pseudocode}
\end{algorithm}

We assume that all particles are of spherical shape and all particle-particle and particle-wall interactions are of Hertzian type. During collision, the contact force can be decomposed into a normal ($\mathbf{f}^n$) and a tangential ($\mathbf{f}^t$) component. The small deformation the particles $i$ and $j$ undergo during collision are approximately modelled as Hertzian overlap ($\xi_{ij}$) (see Fig. \ref{fig:meth_collision}). We then compute the normal force $\mathbf{f}^n_{ij}$ as a function of $\xi_{ij}$ and the material properties of the particle as 

\begin{equation} \label{eqn:Fn}
f^n_{ij} = \max \left\{0,\left[\frac{2}{3} \frac{Y \sqrt{R^\mathrm{eff}_{ij}}}{\left(1-\nu^2\right)}\left(\xi_{ij}^{3 / 2}+A \sqrt{\xi_{ij}} \dot{\xi}_{ij} \right)\right]\right\},
\end{equation}
where, $Y$ is the Young's modulus, $\nu$ is the Poisson ratio, $A$ is the dissipative constant, and
$
R^\mathrm{eff}_{ij} = (\frac{1}{R_i} + \frac{1}{R_j})^{-1}
$, is the effective radius of the interacting particles, $R_i$ and $R_j$ being the radii of particles $i$ and $j$, respectively. The overlap $\xi_{ij}$ is calculated as,

\begin{equation}\label{eqn:deformation}
    \xi_{ij} = R_i + R_j - ||\mathbf{r}_i - \mathbf{r}_j||,
\end{equation}

where, $\mathbf{r}_i$ and $\mathbf{r}_j$ are the position vectors of $i^{\mathrm{th}}$ and $j^{\mathrm{th}}$ particles' center of masses respectively. The rate of change of overlap, $\dot{\xi}_{ij}$, in equation \ref{eqn:Fn} is given as,
\begin{equation}
    \dot{\xi}_{ij} = -\mathbf{v}_{\mathrm{rel},ij} \cdot \hat{\mathbf{e}}_{ij}^n,
\end{equation}
where $\mathbf{v}_\mathrm{rel,ij} = \mathbf{v}_i - \mathbf{v}_j$ is the relative velocity of particles in contact and $\hat{\mathbf{e}}^n$ is the unit normal. In equation \ref{eqn:Fn}, if the material properties ($Y$ and $\nu$) of the interacting particles are different, the effective properties ($Y^{\mathrm{eff}}$, $\nu^{\mathrm{eff}}$) have to be considered. In this study, we assume that all particles and walls possess the same material properties. Note that equation \ref{eqn:Fn} gives the magnitude of the normal force. Normal force vector $\mathbf{f}^n_{ij}$ can be computed as,
\begin{equation} \label{eqn:fn vector}
    \mathbf{f}^n_{ij} = f^n_{ij} \cdot \hat{\mathbf{e}}_{ij}^n.
\end{equation}

During oblique collisions, the interacting particles also experience tangential forces  ($\mathbf{f}^t_{ij}$) because of the surface textures \cite{poschel_computational_2005}. We use the Haff and Werner model \cite{haff_computer_1986} where tangential forces are modeled as a function of tangential relative velocity ($\mathbf{v}^t_{\mathrm{rel},{ij}}$) at the point of contact. 
\begin{equation} \label{eqn:Ft}
    \mathbf{f}^t_{ij} = -\left(\frac{\mathbf{v}^t_{\mathrm{rel},{ij}}}{| \mathbf{v}^t_{\mathrm{rel},{ij}}|}\right)  \mathrm{min}(\gamma^t|\mathbf{v}^t_{\mathrm{rel},{ij}}|, \mu|\mathbf{f}^n_{ij}|),
\end{equation}
where, $\gamma^t$ and $\mu$ are constant viscous damping coefficient and sliding friction coefficient, respectively. The tangential component of the relative velocity, $\mathbf{v}^t_{\mathrm{rel},ij}$, is calculated as
\begin{equation}
    \mathbf{v}^t_{\mathrm{rel},{ij}} = \mathbf{v}_{\mathrm{rel},ij} - (\mathbf{v}_{\mathrm{rel},ij}\cdot \hat{\mathbf{e}}^n_{ij})\hat{\mathbf{e}}^n_{ij} +(R_i\boldsymbol{\omega}_i+R_j\boldsymbol{\omega}_j)\times \hat{\mathbf{e}}^n_{ij}.
\end{equation}
Note that, tangential force in equation \ref{eqn:Ft} is limited by the Coulumb's friction law. The resultant force acting on particle $i$ due to its pairwise interaction with particle $j$ is the sum of normal and tangential forces,
\begin{equation}
    \mathbf{f}_{ij} = 
    \begin{cases}
        \mathbf{f}^n_{ij} + \mathbf{f}^t_{ij},  &\text{   if   } \xi_{ij} > 0\\
        0, &\text{otherwise}.
    \end{cases}
    \label{eqn:fij}
\end{equation}
The tangential force $\mathbf{f}^t_{ij}$ imparts a sliding friction torque $\boldsymbol{\tau}^s_{ij}$ (see Fig. \ref{fig:meth_collision}) which is computed as,
\begin{equation}
    \boldsymbol{\tau}^s_{ij} = \mathbf{f}^t_{ij} \times (-R^{\mathrm{eff}}_{ij}\hat{\mathbf{e}}^n_{ij}).
    \label{eqn:ttij}
\end{equation}
The negative sign in equation \ref{eqn:ttij} arises because the radius vector is computed in the outward radial direction while $\hat{\mathbf{e}}^n_{ij}$ is directed inwards.

When a particle rolls over a surface, in addition to the sliding friction, it experiences hysteresis caused due to micro-sliding, adhesion, shape deformation etc. This imparts an additional torque that resists the rolling motion of the particle. This resistance torque, hereby referred to as the rolling resistance $\boldsymbol{\tau}^r_{ij}$ is computed using Luding's model \cite{luding_cohesive_2008} . We first compute the rolling pseudo-force as,
\begin{equation}
    \mathbf{f}^{r}_{ij} = k^r\boldsymbol{\xi}^r_{ij} - \gamma^{r}\mathbf{v}^r_{ij}, 
    \label{eqn:frij}
\end{equation}
where, $k^r$ is the rolling stiffness, $\gamma^r$ is the viscous damping constant for rolling, $\mathbf{v}^r_{ij}=-R_i(\boldsymbol{\omega}_i-\boldsymbol{\omega}_j)\times \hat{\mathbf{e}}^n_{ij}$ is the relative rolling velocity, and $\boldsymbol{\xi}^r_{ij}$ is the rolling displacement between two time instances $t_1$ and $t_2$ given as 
\begin{equation}
\boldsymbol{\xi}^r_{ij}=\int_{t_1}^{t_2}\mathbf{v}^r_{ij} dt,
\end{equation}
which is numerically approximated for a time duration $\Delta t_{\text{DEM}}$ as $\boldsymbol{\xi}^r_{ij} \approx \mathbf{v}^r_{ij}\Delta t_{\text{DEM}}$.
The `rolling pseudo-force' in equation \ref{eqn:frij} is called so because it doesn't contribute to the total force acting on the either particles \cite{plimpton_lammps_2023}. Further, $\mathbf{f}^{r}_{ij}$ is limited by Coulomb's friction criterion. i.e.,
\begin{equation}
\mathbf{f}^r_{ij} = \mathrm{min}(\mu^rf^n_{ij},||\mathbf{f}^r_{ij}||)\mathbf{k},
\end{equation}
where, $\mathbf{k} = \frac{\mathbf{v}^r_{ij}}{||\mathbf{v}^r_{ij}||}$ is the direction of $\mathbf{f}^r_{ij}$.

The torque due to rolling friction can then be stated as,
\begin{equation}
\boldsymbol{\tau}^r_{ij} = \mathbf{f}^r_{ij} \times (-R^\mathrm{eff}_{ij}\hat{\mathbf{e}}^n_{ij}).
\label{eqn:trij}
\end{equation}

Total torque acting on the particle $i$ because of its interaction with particle $j$ is,
\begin{equation}
    \boldsymbol{\tau}_{ij} = \boldsymbol{\tau}^s_{ij} + \boldsymbol{\tau}^r_{ij}.
    \label{eqn:tij}
\end{equation}

Note that, equations \ref{eqn:fij} and \ref{eqn:tij}, respectively, represent force and torque acting on particle $i$ due to its interactions with particle $j$. Hence, for a system of $N_p$ particles, the total force and torque on particle $i$ is,

\begin{equation} \label{eqn:total force and torque}
\begin{aligned}
    \mathbf{f}_i &= \sum_{j=1, j\neq i}^{N_p} \mathbf{f}_{ij}, \\
    \boldsymbol{\tau}_i &= \sum_{j=1, j\neq i}^{N_p} \boldsymbol{\tau}_{ij}.
\end{aligned}
\end{equation}

Particle-wall interactions can be modeled using the equations similar to \ref{eqn:Fn}-\ref{eqn:tij}. A wall can be seen as a particle of infinite radius. Walls considered in this study are stationary, flat and bear the same material properties as particles. The total force and torque experienced by the $i^\mathrm{th}$ particle because of its collision with walls is also added to the right-hand-side of the respective equations in \ref{eqn:total force and torque}. Lastly, particles' position and translational velocities can be computed from,

\begin{equation} \label{eqn:translational}
    \begin{aligned}
        \frac{d\mathbf{v}_i}{dt} &= \frac{\mathbf{f}_i}{M_i}, \\
        \frac{d\mathbf{r}_i}{dt} &= \mathbf{v}_i,
    \end{aligned}
\end{equation} 
and, the angular position and velocity (assuming spherical particles) can be computed from,
\begin{equation} \label{eqn:angular}
    \begin{aligned}
        \frac{d\boldsymbol{\omega}_i}{dt} &= \frac{\boldsymbol{\tau}_i}{I_i}, \\
        \frac{d\boldsymbol{\theta}_i}{dt} &= \boldsymbol{\omega}_i,
    \end{aligned}
\end{equation}
where, $\boldsymbol{\omega}_i$, $\boldsymbol{\theta}_i$, $M_i$ and $I_i$ are the angular velocity, angular position, mass and moment of inertia of particle $i$ respectively.

We use Gear's integration scheme to solve equations \ref{eqn:translational} and \ref{eqn:angular}. Gear's scheme uses a predictor-corrector approach in which truncated Taylor series expansion is first used to predict the positions, velocities, accelerations etc., of the particles. Forces and torques are then computed using equations \ref{eqn:Fn}-\ref{eqn:total force and torque}. Corrected accelerations, both linear and angular, are calculated from new forces and torques. Lastly, error between corrected and predicted accelerations is used to correct positions and velocities and the process is continued. The algorithm is presented in \ref{alg:pseudocode}.

\subsection{IBM and DEM Coupling} \label{sec:coupling}
While DEM considers a particle as a discrete object (see Fig. \ref{fig:particle in DEM and IBM} (a)), IBM considers it as a collection of Lagrangian markers (see Fig. \ref{fig:particle in DEM and IBM}(b)). All the forces $\mathbf{f}_i$ and torques $\boldsymbol{\tau}_i$ discussed in section \S \ref{sec:dem} are computed for each particle $i$ whereas the collision force discussed in equation \ref{eqn:fb=fc+fcol} refers to the force acting on each Lagrangian marker entailed in a particle. The forces and torques ($\mathbf{f}_i$ and $\boldsymbol{\tau}_i$) calculated from DEM are distributed to each Lagrangian marker point $l$ as $\mathbf{F}_{col}(\mathbf{s}_l,t)$.

The force acting on the $i^\mathrm{th}$ DEM particle, calculated in equation \ref{eqn:total force and torque}, is distributed to each of its constituent Lagrangian marker $l$ as,
\begin{equation}\label{eqn:Ffil}
    \bar{\mathbf{F}}_{i}^l = \frac{\mathbf{f}_i}{n_i},
\end{equation} 
where $n_i$ is the number of Lagrangian markers in the $i^\mathrm{th}$ DEM particle.

The torque acting on each particle is converted into an equivalent system of forces at each marker. If $\boldsymbol{\tau}_i$ is the torque acting on the $i^\mathrm{th}$ particle, then the equivalent force experienced by $l^\mathrm{th}$ marker in that particle is,
\begin{equation}\label{eqn:Fli}
    \tilde{\mathbf{F}}_{i}^l = \frac{|\boldsymbol{\tau}_i|}{\sum_{l=1}^{n_i} ||\mathbf{X}^l_i-\mathbf{r}_i||} \hat{\boldsymbol{\eta}}^l_i,
\end{equation}
where, $\mathbf{X}^l_i$ is the position vector of $l^\mathrm{th}$ marker, $\mathbf{r}_i$ is the position vector of the $i^\mathrm{th}$ particle and $\hat{\boldsymbol{\eta}}$ is the direction of the force which can be computed as,

\begin{equation}\label{eqn:direction of Fli}
    \hat{\boldsymbol{\eta}}^l_i = \frac{\mathbf{X}^l_i \times \hat{\mathbf{a}}_i}{|\mathbf{X}^l_i \times \hat{\mathbf{a}}_i|},
\end{equation}
where, $\hat{\mathbf{a}}_i$ is the axis of rotation of the particle which can be computed using it's angular velocity as,
$$
\hat{\mathbf{a}}_i = \frac{\boldsymbol{\omega}_i}{|\boldsymbol{\omega}_i|}.
$$

 Formulation of equations \ref{eqn:Fli} and \ref{eqn:direction of Fli} can be found in \S \ref{sec:appendixTorqueToForce}.

Finally, the collision force $\mathbf{F}_{col}$ is calculated as,
\begin{equation} \label{eqn:fcol=ffil+ftil}
    \mathbf{F}_{col}(\mathbf{s},t) = \bar{\mathbf{F}}_{i}^l + \tilde{\mathbf{F}}_{i}^l.
\end{equation}

Contact forces evolve on small time scales. To resolve them and to obey the assumption of `small' deformation, DEM demands a significantly smaller timestep compared to IBM. This reduces the timestep size for the entire algorithm and thus increases computational cost. To avoid this, a common practice in CFD-DEM is to use two different timestep sizes, one each for fluid solver and DEM solver. In this study, we keep DEM time step size ($\Delta t_{\mathrm{DEM}}$) one-hundredth of IBM time step size ($\Delta t_{\mathrm{CFD}}$) unless otherwise is mentioned (see algorithm \ref{alg:pseudocode}).

To further reduce computational cost, DEM solver is often coupled with fluid solver once in every few fluid time steps, typically $\thicksim 50-100$ fluid time steps \cite{mondal_coupled_2016}. However, we do not employ this strategy as it comes at the cost of accuracy. We couple the DEM loop with the fluid solver in every fluid time step (see algorithm \ref{alg:pseudocode}).

The IBM methodology explained in \S \ref{sec:cfd-ibm} is available in the well validated IBAMR open-source framework \cite{bhalla_unified_2013}. The DEM method introduced in \S\ref{sec:dem}, together with its coupling to the IBAMR framework described in \S\ref{sec:coupling}, is implemented in an open-source code \cite{sagar_nayak_2025_18107586}. Further development on this code can be tracked in the publicly available GitHub repository \url{https://github.com/nayaksagar/PRELIMPFlow}.

    \begin{figure}[tb]
        \centering
        \includegraphics[width=0.8\linewidth]{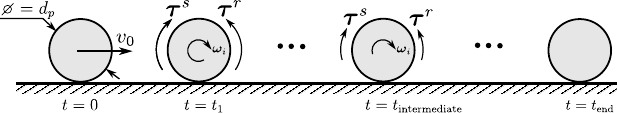}
        \caption{Schematic of a disk rolling on a flat surface}
        \label{fig:rolling_schematic}
    \end{figure}
    \begin{figure}[htb]
        \centering
        \begin{subfigure}[b]{0.45\textwidth}
        \includegraphics[width=\textwidth]{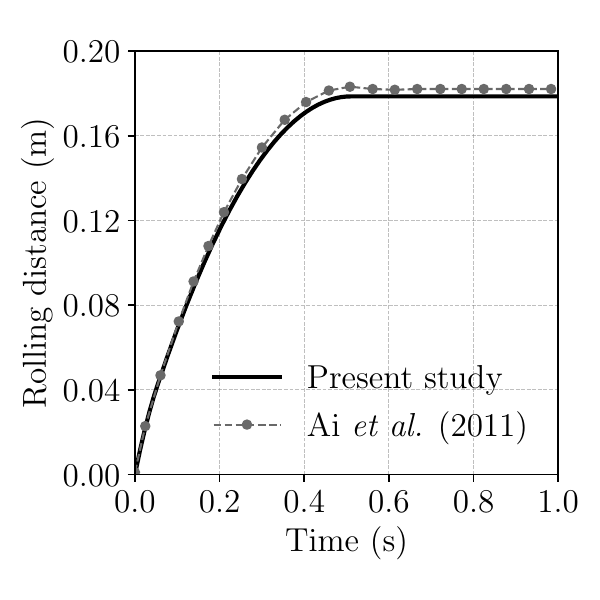}
        \caption{}
        \label{fig:rolling_flat}
      \end{subfigure}
      \hfill
      \begin{subfigure}[b]{0.45\textwidth}
        \includegraphics[width=\textwidth]{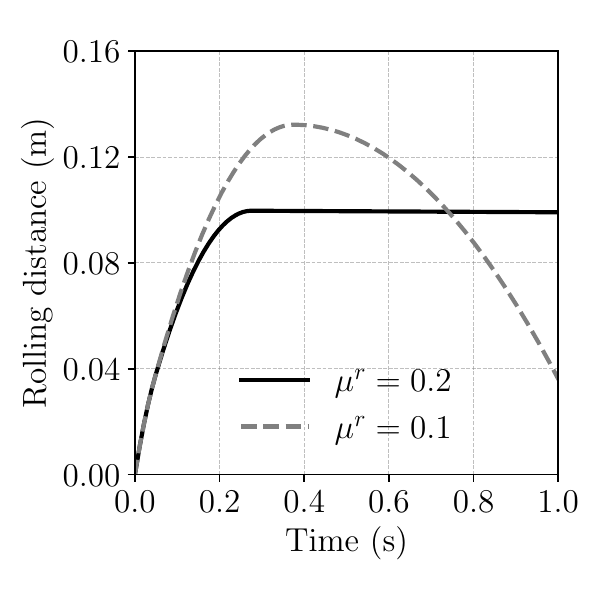}
        \caption{}
        \label{fig:rolling_slope}
      \end{subfigure}
        \caption{Rolling of a circular disk on (a) a flat plane — distance traveled by the disk compared against \cite{ai_assessment_2011}; and (b) an inclined plane — distance traveled by the disk for different rolling friction coefficients}
        \label{fig:rolling_disk_combined_figure}
    \end{figure}

\section{Validation} \label{sec:validAndGridInd}

In this section, we validate the developed particle-resolved IBM–DEM method (described in Sections \ref{sec:dem} and \ref{sec:coupling}) by examining each of the physical effects that play a significant role in pore-clogging phenomena.
To validate the rolling resistance, we simulate the dry-rolling of a disk and a sphere on plane substrates,  in the absence of an ambient fluid in \S \ref{sec:rolling_disk} and \ref{sec:rolling_sphere}. We then study fluid immersed scenarios with particle dynamics: flow through a static packing of particles in \S \ref{sec:packing}, sedimentation of a particle in \S \ref{sec:sedimentation} and drafting-kissing-tumbling of a pair of spherical particles and the clogging of particles near a pore \S \ref{sec:dkt}. 

    \subsection{Dry-rolling of a circular disk} 
    \label{sec:rolling_disk}
    
    We  simulate the rolling of a circular disk on flat and inclined surfaces. These test cases were used in  \cite{ai_assessment_2011} to study different models for rolling friction.
    \subsubsection{On a flat surface}

    A circular disk of diameter $d_p=10mm$ and thickness $t_d=6.67mm$ (Material properties: Density $\rho_s = 1056$ $kg/m^3$, Young's modulus $Y=\SI{40}{MPa}$, Poisson's ratio $\nu=0.49$, Sliding friction coefficient $\mu=0.8$, Rolling friction coefficient $\mu^r = 0.2$) is dropped from a small height (under a constant gravitational acceleration $g=\SI{9.81}{m/s^2}$) onto a flat surface and is allowed to reach equilibrium. Once equilibrium is reached, time is measurement is started (i.e. $t=0$) and the disk's center of mass is given an initial translational velocity of $v_0 = 1 $ m/s in the $x$-direction. A timestep of $\Delta t_{\mathrm{DEM}} = 10^{-5}$ s is used for the simulation. Due to the horizontal sliding friction at the contact between the disk and the flat surface  (equivalently represented by torque $\boldsymbol{\tau}^s$), the disk starts rolling in the $x-y$ plane. 
    Due to the non-point contact of the disk at the substrate, an eccentric normal reaction 
    develops  and equivalently provides a torque $\boldsymbol{\tau}^r$ (as shown in Fig. \ref{fig:rolling_schematic}) that opposes $\boldsymbol{\tau}^s$. As a result, the disk slows down and eventually stops rolling.

    For this case, we run only the DEM loop of the algorithm \ref{alg:pseudocode} and we use the moment of inertia of a disk (i.e. $I=(1/2)MR^2$) in the equation \ref{eqn:angular}. The thickness of disk is such that its volume is equivalent to that of a sphere of the same radius. Figure \ref{fig:rolling_flat} shows the time evolution of the distance traveled by the disk validated against Ai \textit{et al.} \cite{ai_assessment_2011}. 

    \subsubsection{On an inclined plane}
    Analogous to the sliding friction coefficient ($\mu$), the rolling friction coefficient ($\mu^r$) is defined as the tangent of the inclination angle ($\alpha$) of a plane at which the gravitational torque overcomes the rolling resisting torque, causing the disk (or sphere) to roll down. In other words, for any inclination angle $\alpha$, the disk (or sphere) will roll down the slope if, 
    $$
        \mu^r \leq \text{tan}(\alpha).
    $$
    
    To test this, a disk (properties same as in \S \ref{sec:rolling_disk} except $\mu^r$), 
    is dropped from a small height on an inclined plane ($\alpha = \ang{10}$ with $x$-axis) 
    and is allowed to reach equilibrium (under a constant gravitational acceleration $g=\SI{9.8}{m/s^2}$). 
    During the fall and equilibration, any motion in $x$-direction is restricted. 
    Once equilibrium is achieved, an initial translational velocity of $v_0 = \SI{1}{m/s}$ 
     is 
    given to the disk in the direction parallel to the plane. 

    Figure \ref{fig:rolling_slope} shows the time evolution of distance traveled by the disk for different values of rolling friction coefficient ($\mu^r$). At $\mu^r = 0.2$ (i.e. $> \text{tan}(\alpha)$) the disk rolls up the slope, stops and doesn't roll back.  However, at $\mu^r = 0.1$ (i.e. $< \text{tan}(\alpha)$), disk rolls up the slope and rolls back as expected.

    \subsection{Dry-rolling of a sphere on a flat surface} \label{sec:rolling_sphere}
    \begin{figure}
        \centering
        \begin{subfigure}[b]{0.45\textwidth}
        \includegraphics[width=\textwidth]{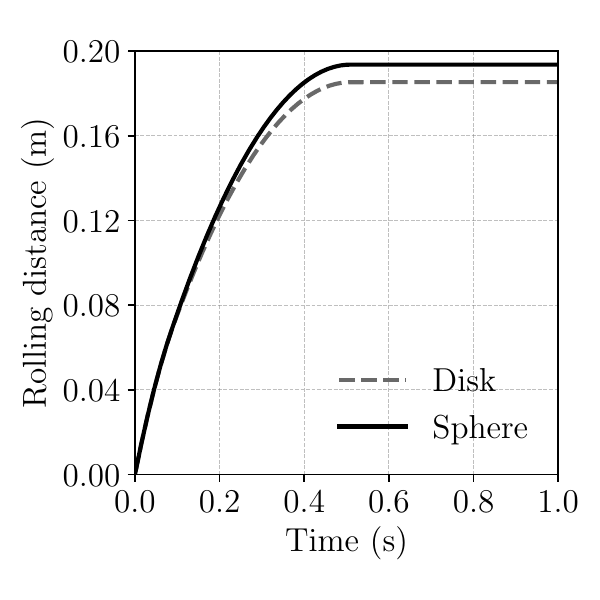}
        \caption{}
        \label{fig:rolling_sphere_distance}
      \end{subfigure}
      \hfill
      \begin{subfigure}[b]{0.45\textwidth}
        \includegraphics[width=\textwidth]{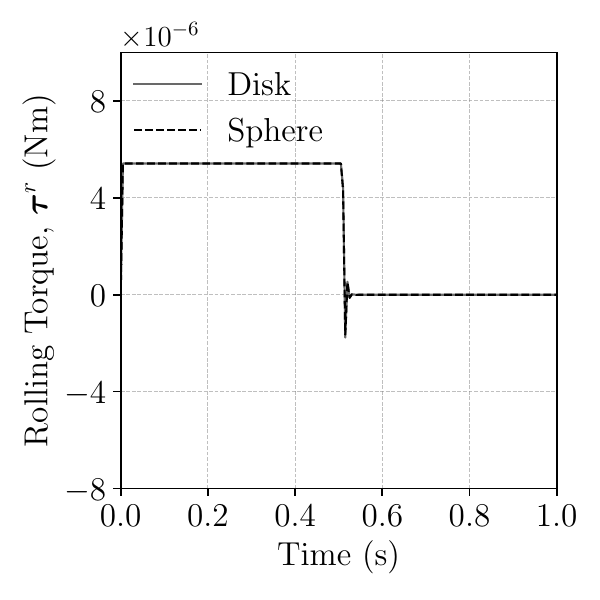}
        \caption{}
        \label{fig:rolling_sphere_torque}
      \end{subfigure}
        \caption{Rolling of a sphere on a flat surface under the influence of rolling friction. Time evolution of (a) distance traveled (b) rolling friction torque experienced by the sphere compared against those experienced by a disk of equal radius and volume.}
        \label{fig:rolling_sphere_combined_figure}
    \end{figure}

    Here we simulate the rolling of a dry sphere on the flat surface and compare the results against those of a disk. The size and material properties of the sphere are same as the disk in section \S \ref{sec:rolling_disk} and the moment of inertia is taken to be $(2/5)MR^2$. Since the sphere and the disks are of the same radius and mass, both experience the same torque as shown in  Fig. \ref{fig:rolling_sphere_torque}. However, having comparatively less moment of inertia, the sphere is expected to travel a longer distance, which is reflected in Fig. \ref{fig:rolling_sphere_distance}.
    
    \subsection{Pressure drop in a particle-packing}
    \label{sec:packing}
    \begin{figure}[tb]
        \centering
        \includegraphics[width=0.8\linewidth]{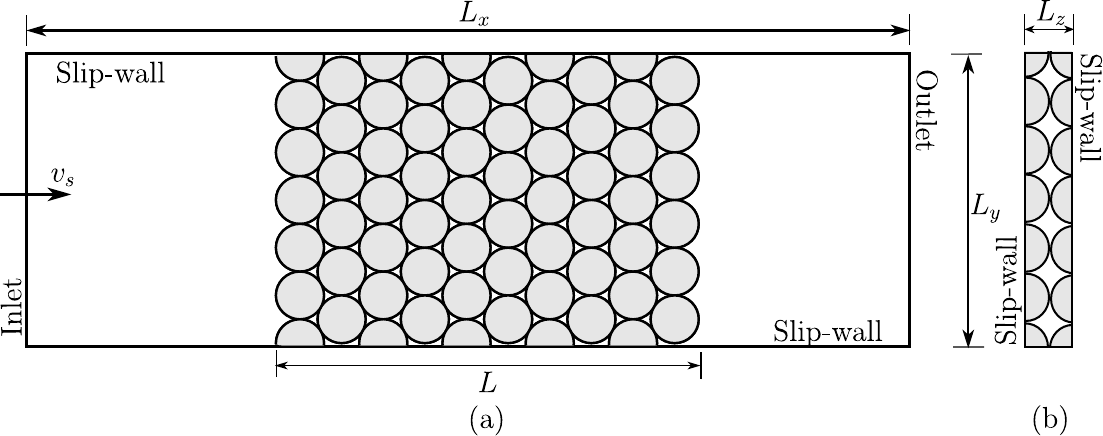}
        \caption{Computational setup for pressure-drop-in-a-particle-packing test. (a) Front view, (b) Side view.}
        \label{fig:ergunSchematic}
    \end{figure}
    \begin{figure}[htb!]
        \centering
        \includegraphics[width=3in]{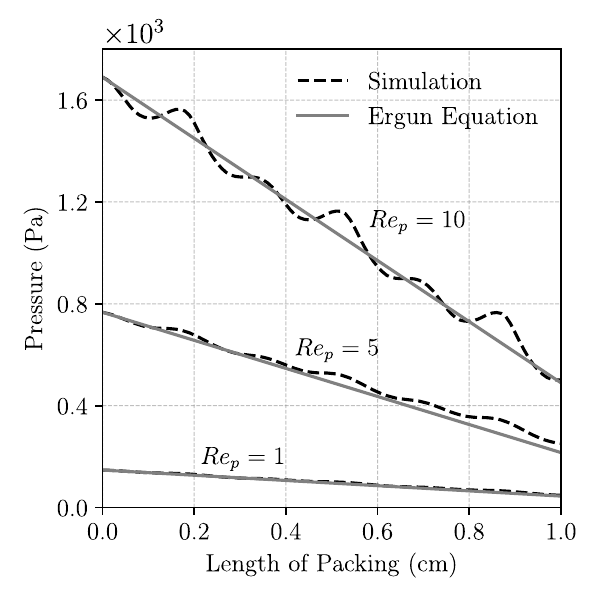}
        \caption{Simulated pressure over the length of the packed bed compared against the Ergun equation \cite{ergun_fluid_1952} for different particle Reynolds numbers $Re_p=\frac{\rho_f v_s d_p}{\mu_f}$.}
        \label{fig:Ergun1_5_10}
    \end{figure}
    When a fluid flows through a porous media like a particle-packing in a clogged system, the resistance offered by the medium causes a pressure drop across it. Ergun \cite{ergun_fluid_1952} proposed a model for the pressure drop per unit length ($L$) of the packing as a function of porosity ($\epsilon$), effective particle diameter ($d_p$) of the constituent particles, fluid properties ($\rho_f$,$\mu_f$), and superficial velocity ($v_s$) as,
    \begin{equation} \label{eqn:ergun}
        \frac{\Delta p}{L} = \frac{150\mu_fv_s}{d_p^2} \frac{(1-\epsilon)^2}{\epsilon^3} + \frac{1.75\rho_fv_s^2}{d_p} \frac{(1-\epsilon)}{\epsilon^3}.
    \end{equation}

    We simulate the flow through a packing of stationary particles. Particles of size $d_p=\SI{0.2}{cm}$ are packed in a $3.6\times 1.2 \times 0.2 \text{ cm}^3$ computational domain as shown in  Fig. \ref{fig:ergunSchematic}. Fluid ($\rho_f=\SI{1000}{kg/m^3}$, $\mu_f=\SI{0.01}{Pa.s}$) enters from the left boundary with a velocity $v_s$, flows through the packed bed and leaves from the right boundary. To minimize the wall-effects, we assign a slip boundary condition to all the other boundaries  and arrange particles such that voids are aligned near the center in the third dimension (see Fig. \ref{fig:ergunSchematic} (b)). 

    The packing results in a porosity of $\epsilon=40.76\%$ and $L=\SI{1.758}{cm}$. The computational domain is discretized into $144\times 48 \times 8$ cells. Pressure is measured on a horizontal line along the length of the packing for different particle Reynolds numbers ($Re_p$) and compared with that predicted from equation \ref{eqn:ergun}. The inlet velocity $v_s$ is varied to achieve different $Re_p$. The resulting plot is shown in Fig. \ref{fig:Ergun1_5_10} which validates that IBM, as implemented in IBAMR is a viable approach to simulate flow through porous media.

    \subsection{Single particle sedimentation}
    \label{sec:sedimentation}
    \begin{table}[tb]
      \centering
      \caption{Parameters for particle sedimentation test.}
      \renewcommand{\arraystretch}{1.3} % increase row height
      \setlength{\tabcolsep}{12pt} % increase column spacing
      \begin{tabular}{c c c c}
        \toprule
        $Re_p$ & $\rho_f \, (\mathrm{kg/m^3})$ & $\mu_f \, (10^{-3}\,\mathrm{Pa.s})$ & $u_\infty \, (\mathrm{m/s})$ \\ \midrule
        4.1    & 965  & 212 & 0.06 \\
        11.6   & 962  & 113 & 0.091 \\
        31.9   & 960  & 58  & 0.128 \\ 
        \bottomrule
      \end{tabular}
      \label{tab:particle_sedimentation}
     \end{table}

    \begin{figure}[htb]
        \centering
        \begin{subfigure}[b]{0.45\textwidth}
        \includegraphics[width=1.13\textwidth]{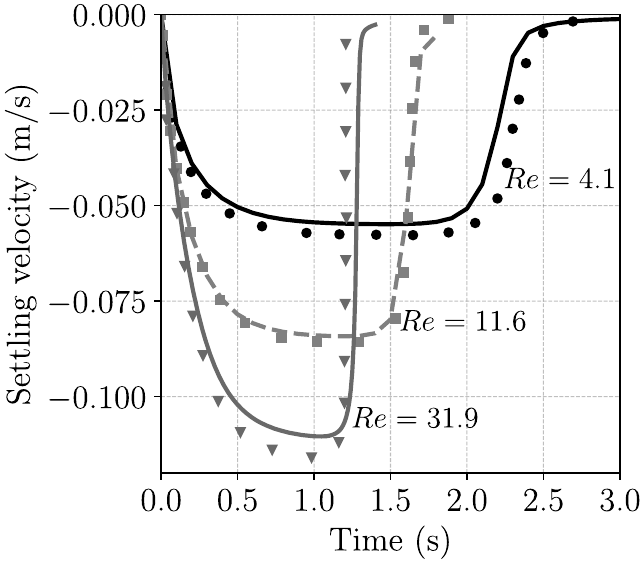}
        \caption{}
        \label{fig:settling_velocity}
      \end{subfigure}
      \hfill
      \begin{subfigure}[b]{0.45\textwidth}
        \includegraphics[width=\textwidth]{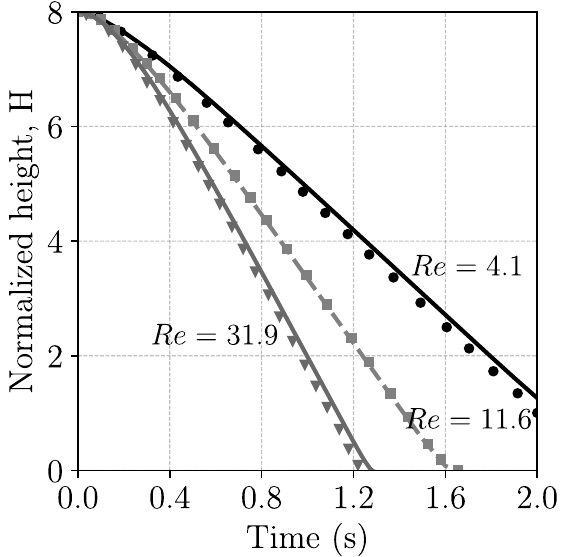}
        \caption{}
        \label{fig:settling_height}
      \end{subfigure}
        \caption{Particle settling in a viscous fluid at different Reynolds numbers. Time evolution of (a) Settling velocity, and (b) Normalized height $\left(H = \frac{h - 0.5d_p}{D_p}\right)$. (Lines: Present simulation, Symbols: Apte \textit{et al.} \cite{apte_numerical_2009})}
        \label{fig:settling_sphere_combined_figure}
    \end{figure}

    Having validated the DEM implementation, we now test the reliability of the coupled IBM-DEM algorithm. The first test along this line is the sedimentation of a spherical particle under gravity in a Newtonian fluid in a closed container.

    A sphere of diameter $d_p = 15mm$ and density $\rho_s = 1125kg/m^3$ is placed at ($5,12,5$) in a computational domain of size $10\times16\times10$ $\mathrm{cm}^3$ and is allowed to settle under gravity ($g=9.81\mathrm{m/s^2}$). A uniform grid of $100\times160\times100$ is used and a fluid timestep of $\Delta t_{\mathrm{CFD}} = 10^{-4}$ is used. Choice of $\Delta t_{\mathrm{DEM}}$ is redundant since no particle-particle or particle-wall interactions are encountered in this test. Following Apte \textit{et al.} \cite{apte_numerical_2009} we repeat this test for three different particle Reynolds numbers ($Re_p = \rho_fu_{\infty} d_p/\mu_f$): 4.1, 11.6 and 31.9. These Reynolds numbers are achieved by varying the fluid density and viscosity as shown in Table \ref{tab:particle_sedimentation}.

    Settling velocity (i.e. velocity in y-direction) and the normalized vertical position ($H=\frac{h-d_p}{d_p}$) of the particle are tracked and plotted against time. Figure \ref{fig:settling_velocity} and Fig. \ref{fig:settling_height} show these plots for different Reynolds numbers (lines). IBM-DEM predictions are in good agreement with \cite{apte_numerical_2009} (symbols).

    \subsection{Drafting-Kissing-Tumbling}
    \label{sec:dkt}
    
    \begin{figure}[htb]
        \centering
        \begin{subfigure}[b]{0.45\textwidth}
        \includegraphics[width=\textwidth]{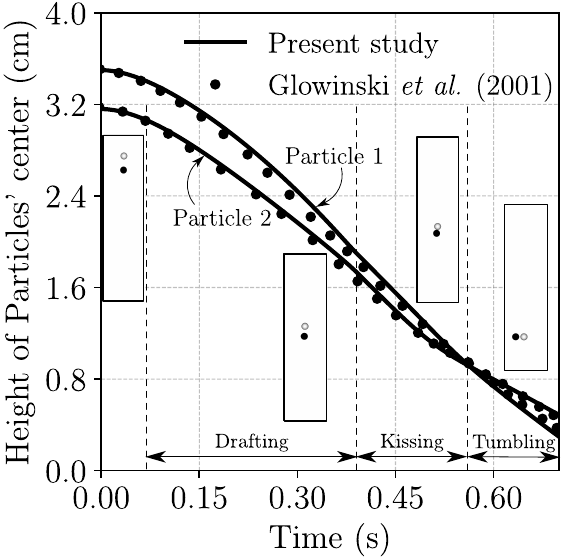}
        \caption{}
        \label{fig:DKT_pos}
      \end{subfigure}
      %\hfill
      \begin{subfigure}[b]{0.45\textwidth}
        \includegraphics[width=\textwidth]{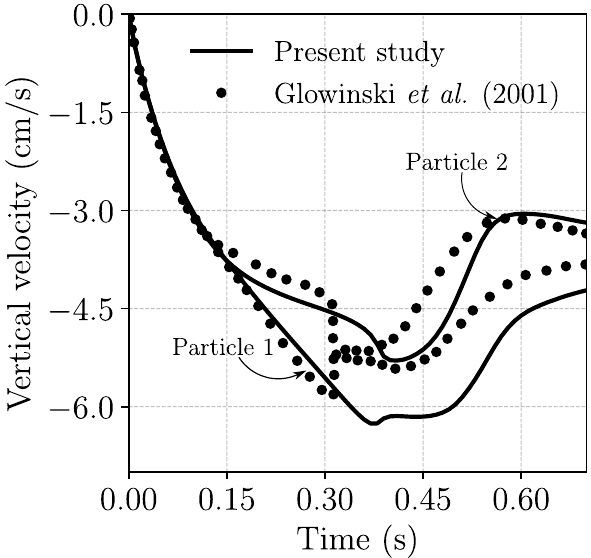}
        \caption{}
        \label{fig:DKT_vel}
      \end{subfigure}
        \caption{Drafting-Kissing-Tumbling phenomenon exhibited by two spherical particles settling in viscous fluid. Time evolution of particles' (a) centers, and (b) vertical velocity compared against Glowinski \textit{et al.}\cite{glowinski_fictitious_2001}}
        \label{fig:dkt_combined_figure}
    \end{figure}
    To further validate the method, we perform the well known `drafting-kissing-tumbling' test. Two spherical particles of same size  ($d_p = 1/6\text{ cm}$) and material, kept initially at a distance vertically, are allowed to settle in a closed container filled with a Newtonian fluid ($\rho_f=1000\mathrm{kg/m^3}, \mu_f=0.001\mathrm{Pa.s}$). In the initial stage, particles only settle under gravity ($9.8\mathrm{m/s^2}$) and hence experience same vertical velocity (see Fig. \ref{fig:DKT_vel}). In the second stage called as `drafting', particle at the top experiences the wake created by the bottom particle. Due to the low pressure in the wake, top particle starts falling with a comparatively greater velocity. This continues till both particles come under contact in a stage called `kissing'. As kissing phase is unstable, particles eventually `tumble' over each other and continue settling.

    For this study we use a $1\times4\times1 \mathrm{cm^3}$ computational domain divided into $60\times240\times60$ uniform grids. We call the top and bottom particles as `Particle 1' and `Particle 2' respectively. At $t=0$ particle 1 is placed at $(0.51,3.5,0.5)$ and particle 2 is placed at $(0.5,3.16,0.5)$. The difference in the x-coordinate of the particles is to induce an instability for tumbling. We take material properties for particles as: Density $\rho_s = 1140 \mathrm{kg/m^3}$, Young's modulus $Y=1\times10^8 \mathrm{MPa}$, Poisson's ratio $\nu=0.3$, Sliding friction coefficient $\mu=0.5$, Rolling friction coefficient $\mu^r = 0.0$. Further, $\Delta t_\mathrm{CFD}=10^{-4}s$ and $\Delta t_\mathrm{DEM}=10^{-7}s$ are chosen. Test is run for a simulated duration of $t=0.8s$.

    Figure \ref{fig:DKT_pos} shows the evolution of height of the particles from the bottom of the domain with time compared with the results of Glowinski \textit{et al.}\cite{glowinski_fictitious_2001}. Both results are in good agreement. Figure \ref{fig:DKT_pos} also shows the extent of different stages. It can be seen, from both plot and the insets, that particles' vertical position flip during the `tumbling' phase. Figure \ref{fig:DKT_vel} shows the time evolution of the vertical velocity of the particles compared against \cite{glowinski_fictitious_2001}. The four distinct stages that particles undergo can be clearly seen in the figure. The initial stage of free-fall is characterized by particles having same velocity ($t=0s-0.16s$). Then Particle 1 is seen gaining acceleration because of getting caught in the low drag region of Particle 2 which is attempting to gain a constant settling velocity ($t=0.16s-0.36s$). At $t\thicksim0.36s$, particles have had contact and as a result Particle 1 looses the momentum and Particle 2 gains it. This can be seen as a small spike and a sharp trough in the respective curves. The velocity variation in the kissing phase i.e.  $t=0.36s-0.57s$ is greatly dependent on the particle-particle interaction model used. This explains the disagreement between our results and Glowinski \textit{et al.}\cite{glowinski_fictitious_2001}. Glowinski \textit{et al.} used a potential force model that acts in the range $R_1+R_2\leq ||\mathbf{r}_1-\mathbf{r}_2|| \leq R_1+R_2+\epsilon$, where $\epsilon$ is a small distance. The fact that it acts before the particles touch each other explains the early appearance of sharp velocity gradients in Glowinski \textit{et al.} compared to the present study ( as seen in Fig. \ref{fig:DKT_vel}). 
    Similar disagreement between DEM and potential-force model can also be observed in 2D studies \cite{kloss_models_2012}.

\section{Results and Discussion}\label{sec:results}
\begin{figure}[tb]
        \centering
        \includegraphics[width=0.8\linewidth]{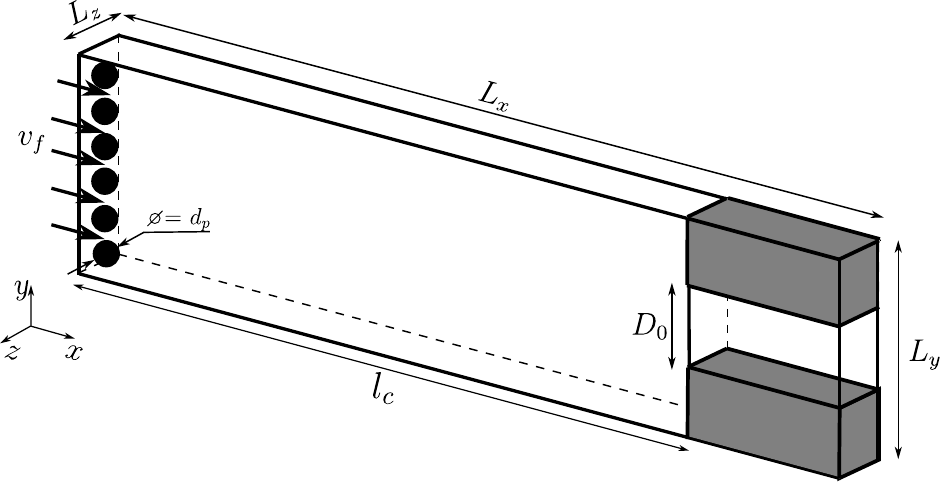}
        \caption{A schematic of the computational setup used for suspension flow through a constricted channel test.}
        \label{fig:mondal_schematic}
    \end{figure}
In this section 
we examine the influence of rolling friction on the bridging behavior in a channel with constriction. 

    \subsection{Suspension flow through a constricted channel} \label{sec:constrictedChannel}

    \begin{table}[htb]
    \caption{Parameters for the suspension flow through constricted channel test.}
    \label{tab:mondal_table}
    \renewcommand{\arraystretch}{1.3} % increase row height
    \setlength{\tabcolsep}{25pt} % increase column spacing
    \begin{center}
    \begin{tabular}{@{}lll@{}}
    \toprule
    \textbf{Parameter}                      & \textbf{Unit}           & \textbf{Value} \\ \midrule
    \multicolumn{3}{c}{\emph{Fluid properties}}                                               \\ \midrule
    Density($\rho_f$)                       & $\mathrm{kg/m^3}$       & 1000           \\
    Viscosity($\mu_f$)                      & $\mathrm{Pa.s}$         & 0.01           \\
    \multicolumn{3}{c}{\emph{Material properties}}                                            \\ \midrule
    Density($\rho_s$)                       & $\mathrm{kg/m^3}$       & 1000           \\
    Young's modulus($Y$)                    & $\mathrm{MPa}$          & 10         \\
    Poisson's ratio($\nu$)                  &                         & 0.23           \\
    Sliding friction coefficient ($\mu$)    &                         & 0.5            \\
    Rolling friction coefficient ($\mu^r$)  &                         & 0.5            \\
    Dissipative constant (A)                &      $\mathrm{s}$                   & 0.001          \\
    Rolling stiffness ($k^r$)               &$\mathrm{N/m}$  & $5\times 10^5$    \\
    Viscous damping for rolling ($\gamma^r$) &$\mathrm{N.s/m^2}$  &2.5 \\
    \bottomrule
    \end{tabular}
\end{center}
    \end{table}

\begin{figure}[htb!]
        \centering
        \includegraphics[width=3in]{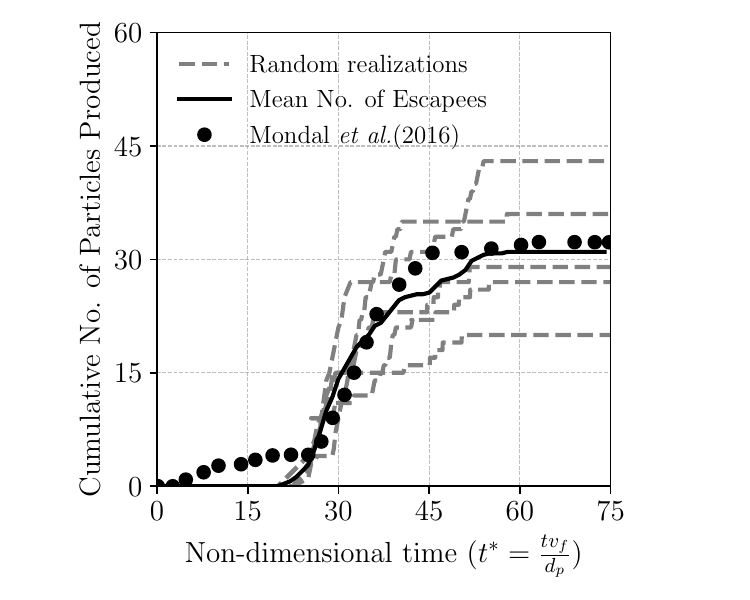}
        \caption{Cumulative number of particles produced versus non-dimensional time.}
        \label{fig:particles_produced}
    \end{figure}
    \begin{figure}[htb]
        \centering
        \includegraphics[width=3in]{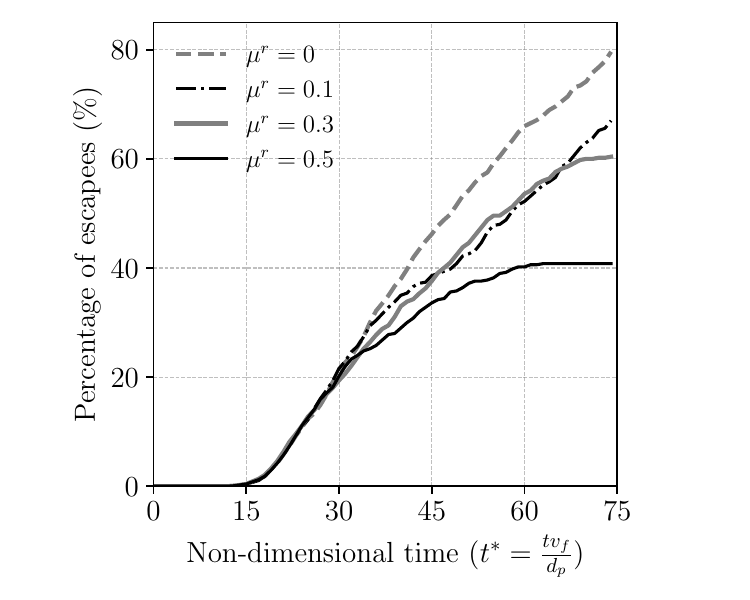}
        \caption{Effect of rolling friction on particle bridging.}
        \label{fig:EffectOfRollingOnClogging}
    \end{figure}

    \begin{figure}[htb]
        \centering
        \includegraphics[width=0.8\linewidth]{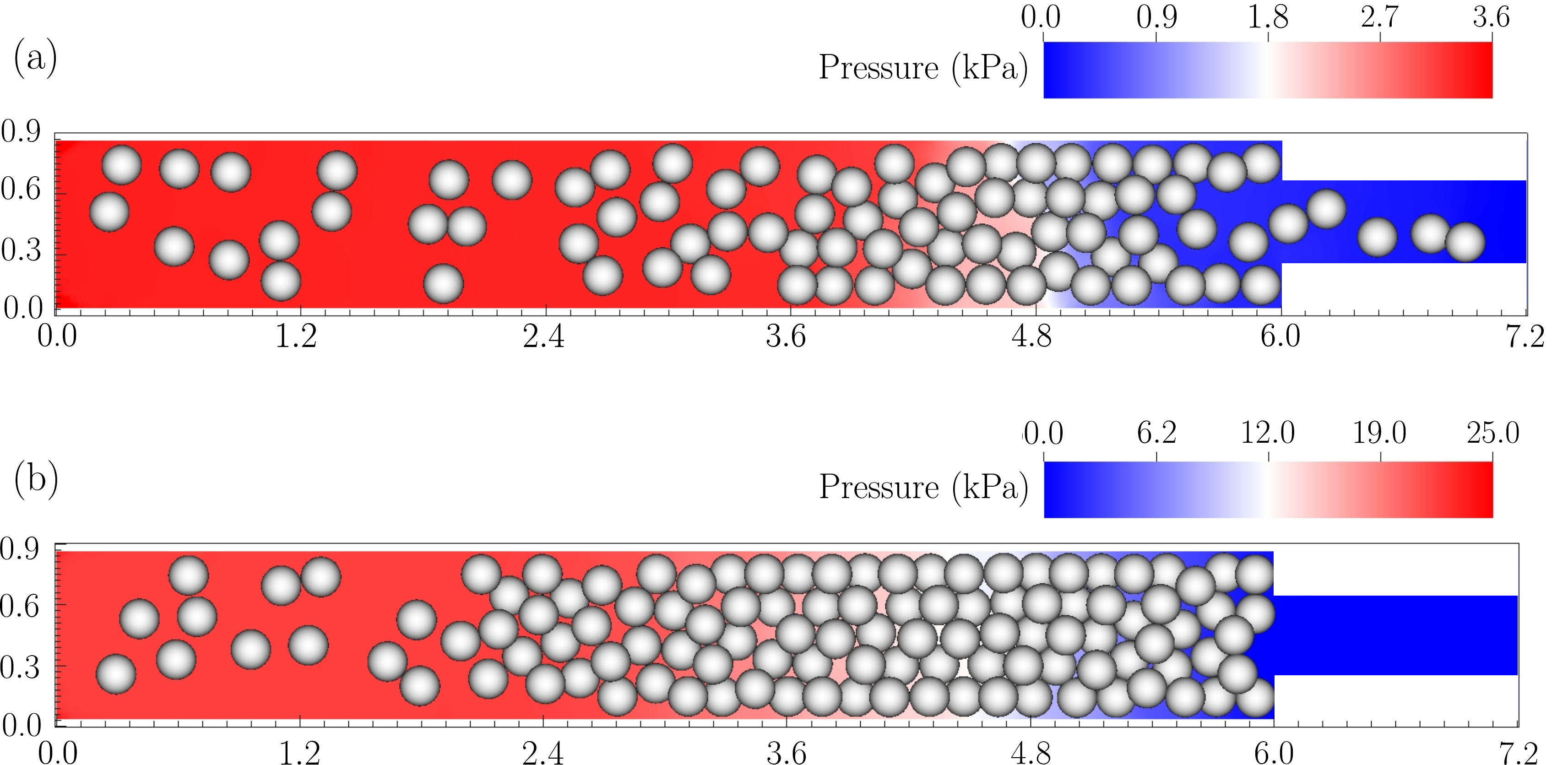}
        \caption{Instantaneous absolute pressure field and particle positions at $t^*=75$ at section $z=\SI{0.15}{cm}$ for (a) $\mu^r=0$ and (b) $\mu^r=0.5$. The larger pressure drop in (b) is due to particle clogging.}
        \label{fig:pressurefield}
    \end{figure}
    
    \begin{figure}[htb]
    \centering
    \includegraphics[width=0.9\linewidth]{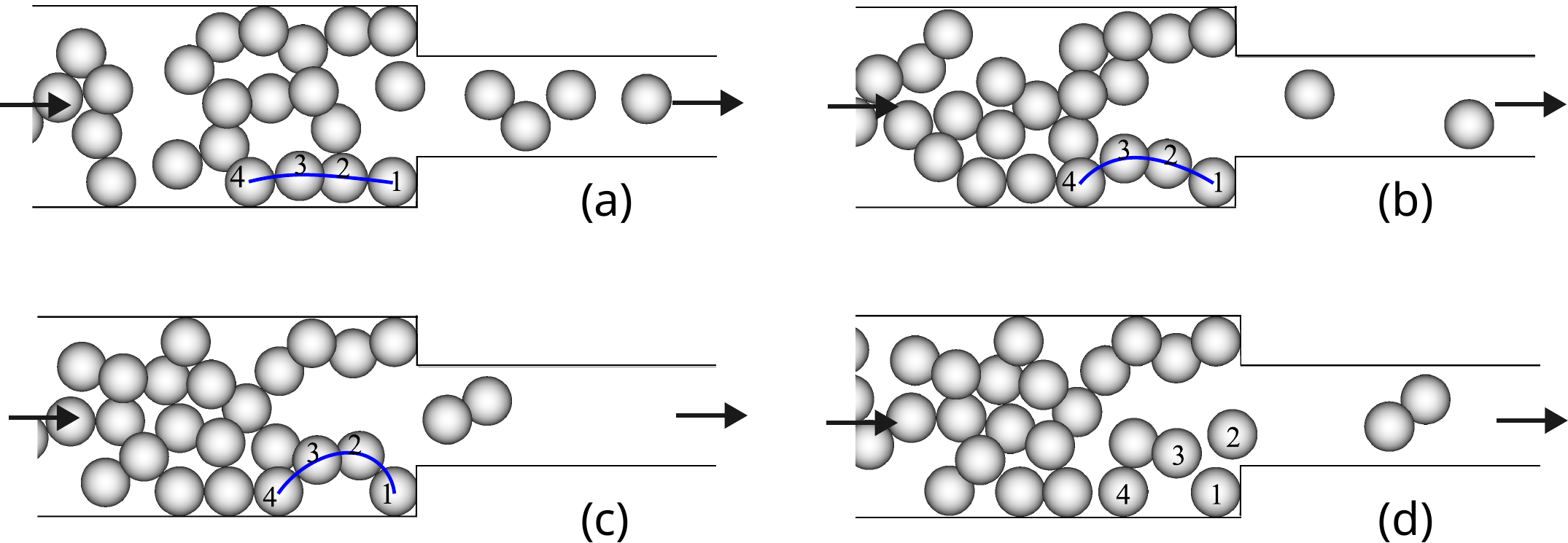}
    \caption{Formation and collapse of particle chain during flow through a constricted channel for the case of $\mu^r=0.5$. Arrows show the direction of the flow.}
    \label{fig:particle_chain}
\end{figure}

    \begin{figure}[htb!]
        \centering
        \includegraphics[width=0.75\linewidth]{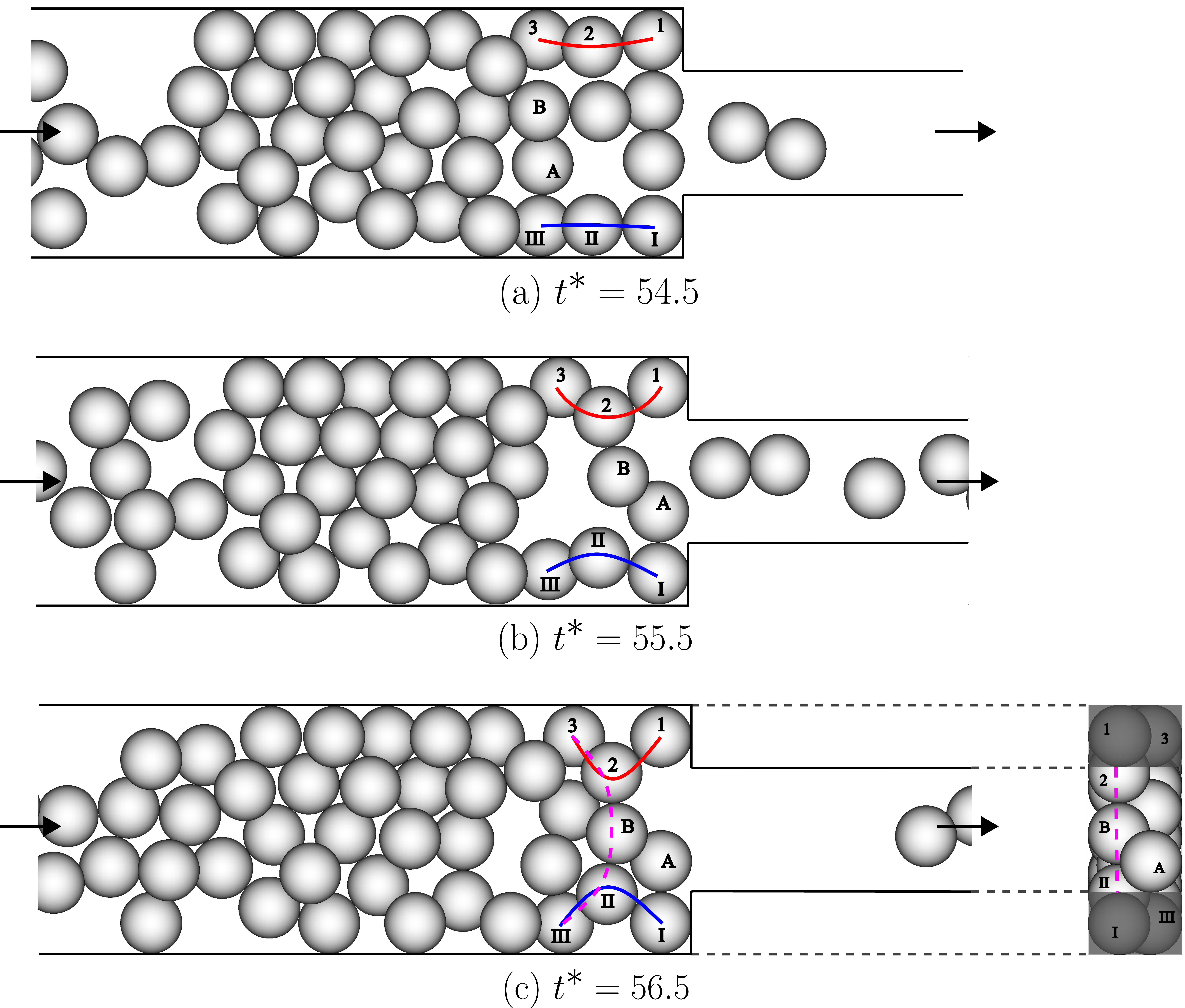}
        \caption{Interaction of free-stream particles (A and B) with particles of convex particle-chain (2 and II) forming a stable bridge for $\mu^r=0.5$. Arrows show the flow direction.}
        \label{fig:twoconvex}
    \end{figure}

    When a suspension is made to flow through a narrow channel, multiple suspended particles simultaneously arrive at the entrance of the channel and form a `bridge' across the entrance creating an obstruction to the flow. The phenomenon of bridging exhibits a stochastic nature and is dependent on parameters like suspension concentration, flow velocity and ratio of particle diameter to channel size \cite{ramachandran_low_2000}.

    Mondal \textit{et al.} \cite{mondal_coupled_2016} performed a numerical study on the effect of various parameters on the bridging phenomenon. To check the reliability of our model in capturing the bridging, we perform a similar study to compare with.
    A computational domain of size $7.2\times1.5\times0.3 \mathrm{cm^3}$ with an orifice of size $D_o$ at a distance $l_c=\SI{6}{cm}$ from the inlet as shown in Fig.  \ref{fig:mondal_schematic} is used for the study.   The domain is discretized into  a uniform grid of resolution $192\times40\times8$ and fluid and DEM times are advanced in steps of $10^{-6}s$ and $10^{-8}s$, respectively. This resolution is higher ($\sim$ 5.33 cells along diameter of each DEM particle) than that used by \cite{mondal_coupled_2016} ($\sim$ 4 cells along the diameter of each DEM particle). An uniform inlet velocity of $v_f = 10\mathrm{cm/s}$ is used at the left face of the domain and pressure-outlet boundary condition is used at the right. All the other boundaries including orifice walls are no-slip walls.

    Particles of similar size $d_p = 0.2\mathrm{cm}$ are introduced along with fluid from the inlet. Following \cite{mondal_coupled_2016}, we define the orifice size ($D_o$) in terms of particle size ($d_p$) using the non-dimensional entity $R_o = D_o/d_p$, and assign the value of $R_o=2.5$ for the current study. Usually, a buffer zone is created near the inlet (outside the actual domain) for the insertion of the particles. However, in this study we keep all the particles inside the computational domain at a distance of one particle radius from the inlet. The $y$ and $z$ coordinate of each particle is randomly chosen. All the particles are inactive at $t=0$ and are activated in batches as the time proceeds. The frequency of particle activation depends on the suspension concentration $\phi$ and flow rate as,
    \begin{equation}
        f = \frac{\phi}{100}\frac{L_yL_zv_f}{V_p},
    \end{equation}
    where, $V_p$ is the volume of each particle, $L_y\times L_z$ is the surface area at the inlet and $\phi$ is the suspension concentration in percentages which is taken to be $20\%$ in this study. Table \ref{tab:mondal_table} shows the fluid and material properties used for the study.

    Particles flow along with the fluid and attempt to enter the orifice simultaneously. As a result, they form a bridge across the orifice and clog the path for other particles. Often, these bridges are unstable and break leading to an outbursts of particles. 
    However, under the right circumstances a stable bridge is formed. We state a particle is subjected to stable bridging if it doesn't reach the outlet within 1.5 times the actual travel time required. We count the number of particles that have reached the outlet without forming a bridge or without being blocked. We then plot number of particles produced against non-dimensional time ($t^* = tv_f/d_p$). Since the phenomenon of particle bridging is stochastic in nature, we repeat the experiment 5 times \cite{he_pore-scale_2025} with randomized initial particle positions at the inlet. 
    %In each trial, particles' positions are initialized using a unique seed to achieve randomization. 
    The mean number of particles produced is then computed and compared with Mondal \textit{et al.}\cite{mondal_coupled_2016} as shown in Fig. \ref{fig:particles_produced}. We found good agreement between both the results. The initial deviation is because in \cite{mondal_coupled_2016} few particles are dispersed inside the domain at $t=0$ which tend to escape early. However, in the present study all the particles are initialized at the inlet.
    
\subsection{Effect of rolling friction} \label{sec:rollingResistance}

    To understand the effect of rolling friction on clogging, we perform simulations of particulate flows through constricted channel similar to that in section \S \ref{sec:constrictedChannel} with varying $\mu^r$. Tests are performed in a channel of height $L_y=\SI{0.8}{cm}$ and dimensionless orifice size $R_o=2$ (other dimensions including particle size are kept same as in section \S \ref{sec:constrictedChannel}). Four different rolling friction coefficients ($\mu^r = 0, 0.1, 0.3,\text{ and } 0.5$) are used for the study. All the other parameters are kept as in table \ref{tab:mondal_table}. In each case, we repeat the test 5 times and record the number of particles that reached the outlet which quantifies the extent of bridging. 
    
    Figure \ref{fig:EffectOfRollingOnClogging} shows the time evolution of mean number of particles collected at the outlet, expressed as a percentage of the total number of particles injected at the inlet for different rolling friction coefficients. At lower rolling friction coefficients we observed that bridging events, if not null, are rare. Bridges formed in these cases are unstable and hence tentative. However, as rolling friction is increased, bridges form more often and are stable. In Fig. \ref{fig:pressurefield} (a), a case with no rolling resistance is shown. The pressure gradient is due to the constriction as well as crowding of the particles near the constriction. However, in Fig. \ref{fig:pressurefield} the pressure build up significantly due to clogging and implies the decrease in permeability due to clogging.  

    Qualitatively, the formation of bridges that lead to clogging can be explained by observing the particle dynamics. Often, the particles get pushed towards bottom or top wall forming a convex chain along the length of the channel as shown in Fig. \ref{fig:particle_chain}(a)-\ref{fig:particle_chain}(c). These particle-chains are unstable and eventually collapse/buckle into the flow stream (see Fig. \ref{fig:particle_chain}(d)). 
    
    At high $\mu^r$, the formation of  particle chains simultaneously at the top and bottom walls become increasingly probable, as shown in Fig. \ref{fig:twoconvex} for a case of $\mu^r=0.5$.
    These convex chains appearing simultaneously above and below the constriction effectively makes the constriction narrower, reducing the flow speed and slowing down the dynamics. This also makes further bridging (across the particle chains) probable, as seen in Fig. \ref{fig:twoconvex}(b). 
    The subsequent bridge formation between the two chains, resulting in a fully clogged scenario is shown in Fig. \ref{fig:twoconvex}(c).  Videos of the simulations with different rolling resistance coefficients showing the above described events leading to clogging are provided as supplementary material with this paper.

\section{Conclusion}\label{sec:conclusion}
Pore clogging results in spatio-temporal permeability deterioration in porous media applications. The mechanical causes of pore clogging has remained less understood owing to difficulties in observing the phenomena using experiments. We developed a IBM-DEM technique that allows for particle and pore resolved flow simulations with an in-house DEM solver. We choose rolling resistance as the parameter of interest to span the non-clogging to clogging regime, and we describe the mechanism of clogging based on our simulations. 

We carefully validate the rolling resistance calculations as well as the flow through packing of particles, through six different test cases of increasing complexity.  
We then reproduce the simulation results of \cite{mondal_coupled_2016}. Using the same system we perform a parametric variation of rolling resistance to qualitatively explain the pore clogging process. The paper is novel for its description on the rolling resistance calculations as well as for studying its effect on an intricate flow phenomena. Since the DEM code is made available with this paper \cite{sagar_nayak_2025_18107586}, the results are easily reproducible. Full three dimensional studies and mapping of the entire parameter space for clogging onset are subjects of a future study using the developed solver.

\section*{Acknowledgments}
This work was supported by resources provided by IIT Delhi HPC facility and The University of Queensland Research Computing Centre’s Bunya supercomputer (\url{https://dx.doi.org/10.48610/wf6c-qy55}), with funding from The University of Queensland, Brisbane, Australia. We acknowledge both institutions for providing the necessary computational resources that enabled this research. We also acknowledge Anusandhan National Research Foundation for their support through the Core Research Grant (CRG) number CRG/2022/008634 and
Startup Research Grant (SRG) number SRG/2022/000436 for computational facility. We also acknowledge the Ministry of Shipping and Inland Waterways, India, for their grant that bought us compute time in IIT Delhi's HPC facilty.

\appendix
\section{Derivation of distribution of Torque as a force on Lagrangian markers}\label{sec:appendixTorqueToForce}
\begin{figure}[htb]
    \centering
    \includegraphics[width=2in]{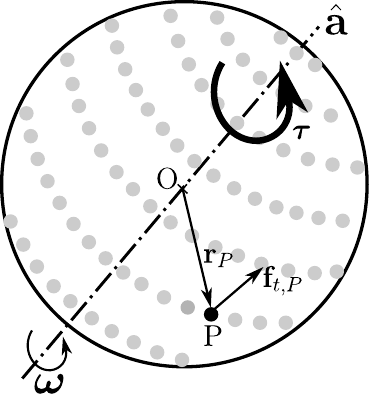}
    \caption{Distribution of torque as forces on constituent Lagrangian markers using equivalent force system.}
    \label{fig:distributionTorqueAsForce}
\end{figure}

Consider a spherical particle constituted of $m$ Lagrangian markers as shown in Fig. \ref{fig:distributionTorqueAsForce}. The total force system on the particle due to its contact with neighbors is split into a moment about the center of mass and a force acting along the center of mass. Let $\boldsymbol{\tau}$ be the torque acting on the particle due to its interaction with other particles or walls. This torque can be equivalently represented by a non-uniform distribution of forces on the Lagrangian markers.

 Let `P' be any Lagrangian marker in the particle and $\mathbf{r}_P$ be its distance vector from the center of the particle $O$.
 %i.e. $\mathbf{r}_P = \mathbf{X}_P-\mathbf{r}$. 
 We construct a force distribution of uniform magnitude  $f_t$, such that the force system is equivalent to the moment about the center of mass O. This force acts in a direction tangential to the axis of rotation of the particle ($\hat{\mathbf{a}}$). The unit vector along this direction is,
 $$
\hat{\boldsymbol{\eta}}_P = \frac{\mathbf{r}_P \times \hat{\mathbf{a}}}{||\mathbf{r}_P \times \hat{\mathbf{a}}||}
 $$
 
Therefore,
 \begin{equation}
      \mathbf{f}_{t,P} = f_t \hat{\boldsymbol{\eta}}_P.
      \label{eqn:ftp}
 \end{equation}
 
The torque created by this force about $\hat{\mathbf{a}}$ is,
$$
\boldsymbol{\tau}_P = \mathbf{r}_P \times \mathbf{f}_{t,P}.
$$

The summation of all such torques on $m$ Lagrangian markers will be equal to the total torque acting on the particle, i.e.,
\begin{equation}
    ||\boldsymbol{\tau}||\hat{\mathbf{a}} = \sum_m \mathbf{r}_{P} \times \mathbf{f}_{t,P}.
    \label{eqn:totalTorque}
\end{equation}

Using equation \ref{eqn:ftp} in \ref{eqn:totalTorque} we get,
$$
    ||\boldsymbol{\tau}||\hat{\mathbf{a}} = f_t\sum_m \mathbf{r}_{P} \times \hat{\boldsymbol{\eta}}_P,
$$

which can be further decomposed as,
$$
||\boldsymbol{\tau}||\hat{\mathbf{a}} = f_t\sum_m ||\mathbf{r}_{P}||(\hat{\mathbf{r}}_{P} \times \hat{\boldsymbol{\eta}}_P),
$$

But $\hat{\mathbf{r}}_{P} \times \hat{\boldsymbol{\eta}}_P = \hat{\mathbf{a}}$. Therefore,
$$
||\boldsymbol{\tau}||\hat{\mathbf{a}} = f_t\sum_m ||\mathbf{r}_{P}|| \hat{\mathbf{a}},
$$

Hence,
\begin{equation}
     f_t = \frac{||\boldsymbol{\tau}||}{\sum_m ||\mathbf{r}_{P}||}.
\end{equation}

\bibliographystyle{elsarticle-num-names} 
\bibliography{bibliography}

@Article{balevicius_effect_2012,
  author   = {Balevičius, R. and Sielamowicz, I. and Mróz, Z. and Kačianauskas, R.},
  title    = {Effect of rolling friction on wall pressure, discharge velocity and outflow of granular material from a flat-bottomed bin},
  journal  = {Particuology},
  year     = {2012},
  volume   = {10},
  number   = {6},
  pages    = {672--682},
  month    = dec,
  issn     = {16742001},
  language = {en},
  urldate  = {2025-10-15},
}

@Article{kloss_models_2012,
  author   = {Kloss, Christoph and Goniva, Christoph and Hager, Alice and Amberger, Stefan and Pirker, Stefan},
  title    = {Models, algorithms and validation for opensource {DEM} and {CFD}-{DEM}},
  journal  = {Prog. Comput. Fluid Dyn. An Int. J.},
  year     = {2012},
  volume   = {12},
  number   = {2/3},
  pages    = {140},
  issn     = {1468-4349, 1741-5233},
  language = {en},
  urldate  = {2022-09-13},
}

@Article{li_coupled_2020,
  author   = {Li, Jia and Qiu, Zhengsong and Zhong, Hanyi and Zhao, Xin and Huang, Weian},
  title    = {Coupled {CFD}-{DEM} analysis of parameters on bridging in the fracture during lost circulation},
  journal  = {J. Petroleum Sci. Eng.},
  year     = {2020},
  volume   = {184},
  pages    = {106501},
  month    = jan,
  issn     = {0920-4105},
  urldate  = {2025-10-10},
}

@Article{di_vaira_hydrodynamic_2023,
  author   = {Di Vaira, Nathan J. and {\L}aniewski-Wo{\l}{\l}k, {\L}ukasz and Johnson, Raymond L. and Aminossadati, Saiied M. and Leonardi, Christopher R.},
  title    = {Hydrodynamic clogging of micro-particles in planar channels under electrostatic forces},
  journal  = {J. Fluid Mech.},
  year     = {2023},
  volume   = {960},
  pages    = {A34},
  month    = apr,
  issn     = {0022-1120, 1469-7645},
  language = {en},
  urldate  = {2025-10-10},
}

@Article{zheng_interparticle_2024,
  author   = {Zheng, Q. J. and Yang, R. Y. and Zeng, Q. H. and Zhu, H. P. and Dong, K. J. and Yu, A. B.},
  title    = {Interparticle forces and their effects in particulate systems},
  journal  = {Powder Technol.},
  year     = {2024},
  volume   = {436},
  pages    = {119445},
  month    = mar,
  issn     = {0032-5910},
  urldate  = {2025-10-10},
}

@Article{elrahmani_clogging_2023,
  author     = {Elrahmani, Ahmed and Al-Raoush, Riyadh I. and Seers, Thomas D.},
  title      = {Clogging and permeability reduction dynamics in porous media: {A} numerical simulation study},
  journal    = {Powder Technol.},
  year       = {2023},
  volume     = {427},
  pages      = {118736},
  month      = sep,
  issn       = {0032-5910},
  shorttitle = {Clogging and permeability reduction dynamics in porous media},
  urldate    = {2025-05-22},
}

@Article{samari_kermani_direct_2020,
  author     = {Samari Kermani, Mandana and Jafari, Saeed and Rahnama, Mohammad and Raoof, Amir},
  title      = {Direct pore scale numerical simulation of colloid transport and retention. {Part} {I}: {Fluid} flow velocity, colloid size, and pore structure effects},
  journal    = {Adv. Water Resour.},
  year       = {2020},
  volume     = {144},
  pages      = {103694},
  month      = oct,
  issn       = {03091708},
  language   = {en},
  shorttitle = {Direct pore scale numerical simulation of colloid transport and retention. {Part} {I}},
  urldate    = {2024-04-19},
}

@Article{lohaus_what_2018,
  author   = {Lohaus, J. and Perez, Y. M. and Wessling, M.},
  title    = {What are the microscopic events of colloidal membrane fouling?},
  journal  = {J. Membr. Sci.},
  year     = {2018},
  volume   = {553},
  pages    = {90--98},
  month    = may,
  issn     = {0376-7388},
  urldate  = {2025-10-10},
}

@Article{mirabolghasemi_prediction_2015,
  author   = {Mirabolghasemi, Maryam and Prodanović, Maša and DiCarlo, David and Ji, Hongyu},
  title    = {Prediction of empirical properties using direct pore-scale simulation of straining through {3D} microtomography images of porous media},
  journal  = {J. Hydrol.},
  year     = {2015},
  volume   = {529},
  pages    = {768--778},
  month    = oct,
  issn     = {00221694},
  language = {en},
  urldate  = {2025-10-10},
}

@Article{tsuji_discrete_1993,
  author   = {Tsuji, Y. and Kawaguchi, T. and Tanaka, T.},
  title    = {Discrete particle simulation of two-dimensional fluidized bed},
  journal  = {Powder Technol.},
  year     = {1993},
  volume   = {77},
  number   = {1},
  pages    = {79--87},
  month    = oct,
  issn     = {0032-5910},
  urldate  = {2025-10-10},
}

@Article{krueger_overview_1986,
  author   = {Krueger, Roland F.},
  title    = {An overview of formation damage and well productivity in oilfield operations},
  journal  = {J. Petroleum Technol.},
  year     = {1986},
  volume   = {38},
  number   = {02},
  pages    = {131--152},
  month    = feb,
  issn     = {0149-2136, 1944-978X},
  language = {en},
  urldate  = {2025-01-14},
}

@Article{yang_stochastic_2020,
  author   = {Yang, Yulong and Yuan, Weifeng and Hou, Jirui and You, Zhenjiang and Li, Jun and Liu, Yang},
  title    = {Stochastic and upscaled analytical modeling of fines migration in porous media induced by low-salinity water injection},
  journal  = {Appl. Math. Mech.},
  year     = {2020},
  volume   = {41},
  number   = {3},
  pages    = {491--506},
  month    = mar,
  issn     = {0253-4827, 1573-2754},
  language = {en},
  urldate  = {2023-09-08},
}

@Article{you_fines_2019,
  author     = {You, Z. and Badalyan, A. and Yang, Y. and Bedrikovetsky, P. and Hand, M.},
  title      = {Fines migration in geothermal reservoirs: {Laboratory} and mathematical modelling},
  journal    = {Geothermics},
  year       = {2019},
  volume     = {77},
  pages      = {344--367},
  month      = jan,
  issn       = {03756505},
  language   = {en},
  shorttitle = {Fines migration in geothermal reservoirs},
  urldate    = {2023-09-08},
}

@Article{you_mathematical_2016,
  author   = {You, Zhenjiang and Yang, Yulong and Badalyan, Alexander and Bedrikovetsky, Pavel and Hand, Martin},
  title    = {Mathematical modelling of fines migration in geothermal reservoirs},
  journal  = {Geothermics},
  year     = {2016},
  volume   = {59},
  pages    = {123--133},
  month    = jan,
  issn     = {03756505},
  language = {en},
  urldate  = {2023-09-08},
}

@Article{yuan_comprehensive_2018,
  author   = {Yuan, Bin and Wood, David A.},
  title    = {A comprehensive review of formation damage during enhanced oil recovery},
  journal  = {J. Petroleum Sci. Eng.},
  year     = {2018},
  volume   = {167},
  pages    = {287--299},
  month    = aug,
  issn     = {0920-4105},
  urldate  = {2025-05-22},
}

@Article{he_pore-scale_2025,
  author   = {He, Haiyang and Xiong, Xiaofeng and Wu, Ting and Hu, Ran and Chen, Yi-Feng and Yang, Zhibing},
  title    = {Pore-scale study of particle transport and clogging mechanisms in a porous micromodel},
  journal  = {Sep. Purif. Technol.},
  year     = {2025},
  volume   = {362},
  pages    = {131929},
  month    = jul,
  issn     = {1383-5866},
  urldate  = {2025-10-10},
}

@Article{ramachandran_plugging_1999,
  author   = {Ramachandran, Venkatachalam and Fogler, H. Scott},
  title    = {Plugging by hydrodynamic bridging during flow of stable colloidal particles within cylindrical pores},
  journal  = {J. Fluid Mech.},
  year     = {1999},
  volume   = {385},
  pages    = {129--156},
  month    = apr,
  issn     = {1469-7645, 0022-1120},
  language = {en},
  urldate  = {2025-10-10},
}

@Article{vani_role_2024,
  author   = {Vani, Nathan and Escudier, Sacha and Jeong, Deok-Hoon and Sauret, Alban},
  title    = {Role of the constriction angle on the clogging by bridging of suspensions of particles},
  journal  = {Phys. Review Research},
  year     = {2024},
  volume   = {6},
  number   = {3},
  pages    = {L032060},
  month    = sep,
  issn     = {2643-1564},
  language = {en},
  urldate  = {2024-09-16},
}

@Article{ergun_fluid_1952,
  author  = {Ergun, S.},
  title   = {Fluid flow through packed columns},
  journal = {Chem. Eng. Prog.},
  year    = {1952},
  volume  = {48},
  number  = {2},
  pages   = {89},
  urldate = {2025-09-22},
}

@Article{haff_computer_1986,
  author   = {Haff, P. K. and Werner, B. T.},
  title    = {Computer simulation of the mechanical sorting of grains},
  journal  = {Powder Technol.},
  year     = {1986},
  volume   = {48},
  number   = {3},
  pages    = {239--245},
  month    = nov,
  issn     = {0032-5910},
  urldate  = {2025-08-26},
}

@Article{glowinski_distributed_1999,
  author   = {Glowinski, R.},
  title    = {A distributed lagrange multiplier/fictitious domain method for particulate flows},
  journal  = {Int. J. Multiphase Flow},
  year     = {1999},
  language = {en},
}

@Article{luding_cohesive_2008,
  author     = {Luding, Stefan},
  title      = {Cohesive, frictional powders: contact models for tension},
  journal    = {Granul. Matter},
  year       = {2008},
  volume     = {10},
  number     = {4},
  pages      = {235--246},
  month      = jun,
  issn       = {1434-7636},
  language   = {en},
  shorttitle = {Cohesive, frictional powders},
  urldate    = {2025-08-25},
}

@Misc{plimpton_lammps_2023,
  author     = {Plimpton, Steven J. and Kohlmeyer, Axel and Thompson, Aidan P. and Moore, Stan G. and Berger, Richard},
  title      = {{LAMMPS}: {Large}-scale {Atomic}/{Molecular} {Massively} {Parallel} {Simulator}},
  month      = aug,
  year       = {2023},
  publisher  = {Zenodo},
  shorttitle = {{LAMMPS}},
  urldate    = {2025-08-25},
}

@Article{ramachandran_low_2000,
  author    = {Ramachandran, Venkatachalam and Venkatesan, Ramachandran and Tryggvason, Grétar and Scott Fogler, H.},
  title     = {Low {Reynolds} number interactions between colloidal particles near the entrance to a cylindrical pore},
  journal   = {J. Colloid Interface Sci.},
  year      = {2000},
  volume    = {229},
  number    = {2},
  pages     = {311--322},
  month     = sep,
  issn      = {00219797},
  copyright = {https://www.elsevier.com/tdm/userlicense/1.0/},
  language  = {en},
  urldate   = {2025-08-13},
}

@Article{bhalla_unified_2013,
  author   = {Bhalla, Amneet Pal Singh and Bale, Rahul and Griffith, Boyce E. and Patankar, Neelesh A.},
  title    = {A unified mathematical framework and an adaptive numerical method for fluid–structure interaction with rigid, deforming, and elastic bodies},
  journal  = {J. Comput. Phys.},
  year     = {2013},
  volume   = {250},
  pages    = {446--476},
  month    = oct,
  issn     = {00219991},
  language = {en},
  urldate  = {2023-08-02},
}

@Article{sharma_coupled_2022,
  author   = {Sharma, Govind and Nangia, Nishant and Bhalla, Amneet Pal Singh and Ray, Bahni},
  title    = {A coupled distributed {Lagrange} multiplier ({DLM}) and discrete element method ({DEM}) approach to simulate particulate flow with collisions},
  journal  = {Powder Technol.},
  year     = {2022},
  volume   = {398},
  pages    = {117091},
  month    = jan,
  issn     = {00325910},
  language = {en},
  urldate  = {2023-08-02},
}

@Misc{griffith_adaptive_2018,
  author    = {Griffith, Boyce},
  title     = {An adaptive and distributed-memory parallel implementation of the immersed boundary ({IB}) method},
  year      = {2018},
  publisher = {IBAMR},
  urldate   = {2025-08-13},
}

@Article{shirgaonkar_new_2009,
  author   = {Shirgaonkar, Anup A. and MacIver, Malcolm A. and Patankar, Neelesh A.},
  title    = {A new mathematical formulation and fast algorithm for fully resolved simulation of self-propulsion},
  journal  = {J. Comput. Phys.},
  year     = {2009},
  volume   = {228},
  number   = {7},
  pages    = {2366--2390},
  month    = apr,
  issn     = {0021-9991},
  urldate  = {2025-08-13},
}

@Article{liu_mechanisms_2024,
  author   = {Liu, Shitao and Shikhov, Igor and Arns, Christoph},
  title    = {Mechanisms of pore-clogging using a high-resolution {CFD}-{DEM} colloid transport model},
  journal  = {Transp. Porous Media},
  year     = {2024},
  volume   = {151},
  number   = {4},
  pages    = {831--851},
  month    = mar,
  issn     = {0169-3913, 1573-1634},
  language = {en},
  urldate  = {2024-04-15},
}

@Article{glowinski_fictitious_2001,
  author     = {Glowinski, R. and Pan, T. W. and Hesla, T. I. and Joseph, D. D. and Périaux, J.},
  title      = {A fictitious domain approach to the direct numerical simulation of incompressible viscous flow past moving rigid bodies: {Application} to particulate flow},
  journal    = {J. Comput. Phys.},
  year       = {2001},
  volume     = {169},
  number     = {2},
  pages      = {363--426},
  month      = may,
  issn       = {0021-9991},
  shorttitle = {A {Fictitious} {Domain} {Approach} to the {Direct} {Numerical} {Simulation} of {Incompressible} {Viscous} {Flow} past {Moving} {Rigid} {Bodies}},
  urldate    = {2025-06-10},
}

@Article{apte_numerical_2009,
  author   = {Apte, Sourabh V. and Martin, Mathieu and Patankar, Neelesh A.},
  title    = {A numerical method for fully resolved simulation ({FRS}) of rigid particle–flow interactions in complex flows},
  journal  = {J. Comput. Phys.},
  year     = {2009},
  volume   = {228},
  number   = {8},
  pages    = {2712--2738},
  month    = may,
  issn     = {0021-9991},
  urldate  = {2025-06-07},
}

@Article{ai_assessment_2011,
  author   = {Ai, Jun and Chen, Jian-Fei and Rotter, J. Michael and Ooi, Jin Y.},
  title    = {Assessment of rolling resistance models in discrete element simulations},
  journal  = {Powder Technol.},
  year     = {2011},
  volume   = {206},
  number   = {3},
  pages    = {269--282},
  month    = jan,
  issn     = {0032-5910},
  urldate  = {2024-12-17},
}

@Book{poschel_computational_2005,
  title      = {Computational {Granular} {Dynamics}: {Models} and {Algorithms}},
  publisher  = {Springer-Verlag},
  year       = {2005},
  author     = {Pöschel, Thorsten and Schwager, Thomas},
  address    = {Berlin ; New York},
  isbn       = {978-3-540-21485-4},
  language   = {en},
  shorttitle = {Computational granular dynamics},
}

@Article{mondal_coupled_2016,
  author   = {Mondal, Somnath and Wu, Chu-Hsiang and Sharma, Mukul M.},
  title    = {Coupled {CFD}-{DEM} simulation of hydrodynamic bridging at constrictions},
  journal  = {Int. J. Multiphase Flow},
  year     = {2016},
  volume   = {84},
  pages    = {245--263},
  month    = sep,
  issn     = {03019322},
  language = {en},
  urldate  = {2024-06-11},
}

@Article{herzig_flow_1970,
  author  = {Herzig, J. P. and Leclerc, D. M. and Goff, P. Le.},
  title   = {Flow of suspensions through porous media—{Application} to deep filtration},
  journal = {Ind. \& Eng. Chem.},
  year    = {1970},
  volume  = {62},
  number  = {5},
  pages   = {8--35},
  month   = may,
  issn    = {0019-7866},
  urldate = {2024-01-04},
}

@Article{chalk_pore_2012,
  author   = {Chalk, P. and Gooding, N. and Hutten, S. and You, Z. and Bedrikovetsky, P.},
  title    = {Pore size distribution from challenge coreflood testing by colloidal flow},
  journal  = {Chem. Eng. Res. Des.},
  year     = {2012},
  volume   = {90},
  number   = {1},
  pages    = {63--77},
  month    = jan,
  issn     = {0263-8762},
  language = {en},
  urldate  = {2023-06-06},
}

@Book{civan2023reservoir,
  title     = {Reservoir {Formation} {Damage}: {Fundamentals}, {Modeling}, {Assessment}, and {Mitigation}},
  publisher = {Gulf Professional Publishing},
  year      = {2023},
  author    = {Civan, Faruk},
}

@Article{katz1987prediction,
  author    = {Katz, A. J. and Thompson, A. H.},
  title     = {Prediction of rock electrical conductivity from mercury injection measurements},
  journal   = {J. Geophys. Res. Solid Earth},
  year      = {1987},
  volume    = {92},
  number    = {B1},
  pages     = {599--607},
  publisher = {Wiley Online Library},
}

@Article{sendekie2016colloidal,
  author    = {Sendekie, Zenamarkos B. and Bacchin, Patrice},
  title     = {Colloidal jamming dynamics in microchannel bottlenecks},
  journal   = {Langmuir},
  year      = {2016},
  volume    = {32},
  number    = {6},
  pages     = {1478--1488},
  publisher = {ACS Publications},
}

@Article{jegatheesan2005deep,
  author    = {Jegatheesan, Veeriah and Vigneswaran, S.},
  title     = {Deep bed filtration: mathematical models and observations},
  journal   = {Crit. Rev. Environ. Sci. Technol.},
  year      = {2005},
  volume    = {35},
  number    = {6},
  pages     = {515--569},
  publisher = {Taylor \& Francis},
}

@Article{nelson2011new,
  author    = {Nelson, Kirk E. and Ginn, Timothy R.},
  title     = {New collector efficiency equation for colloid filtration in both natural and engineered flow conditions},
  journal   = {Water Resour. Res.},
  year      = {2011},
  volume    = {47},
  number    = {5},
  publisher = {Wiley Online Library},
}

@Article{cundall1979discrete,
  author    = {Cundall, Peter A. and Strack, Otto D. L.},
  title     = {A discrete numerical model for granular assemblies},
  journal   = {Geotechnique},
  year      = {1979},
  volume    = {29},
  number    = {1},
  pages     = {47--65},
  publisher = {Thomas Telford Ltd},
}

@Article{goniva2012influence,
  author    = {Goniva, Christoph and Kloss, Christoph and Deen, Niels G. and Kuipers, Johannes A. M. and Pirker, Stefan},
  title     = {Influence of rolling friction on single spout fluidized bed simulation},
  journal   = {Particuology},
  year      = {2012},
  volume    = {10},
  number    = {5},
  pages     = {582--591},
  publisher = {Elsevier},
}

@Article{shahzad2018aggregation,
  author    = {Shahzad, Khurram and Aeken, Wouter Van and Mottaghi, Milad and Kamyab, Vahid Kazemi and Kuhn, Simon},
  title     = {Aggregation and clogging phenomena of rigid microparticles in microfluidics: comparison of a discrete element method ({DEM}) and {CFD}--{DEM} coupling method},
  journal   = {Microfluid. Nanofluid.},
  year      = {2018},
  volume    = {22},
  number    = {9},
  pages     = {104},
  publisher = {Springer},
}

@Article{wang2023unresolved,
  author    = {Wang, Peng and Li, Qingyang and Ma, Tengfei and Ou, Xuelian and Shen, Yanxin and Yang, Yue and Tian, Xiaofeng},
  title     = {An unresolved cfd-dem method for studying migration and clogging of fine particles through a packed bed},
  journal   = {Ironmak. \& Steelmak.},
  year      = {2023},
  volume    = {50},
  number    = {11},
  pages     = {1618--1630},
  publisher = {SAGE Publications Sage UK: London, England},
}

@Software{sagar_nayak_2025_18107586,
  author    = {Nayak, Sagar},
  title     = {nayaksagar/{PRELIMPFlow}: {Initial} public release},
  month     = dec,
  year      = {2025},
  publisher = {Zenodo},
  swhid     = {swh:1:dir:6211e9022d2ae33e4a31ce40236108fce6f80a77 ;origin=https://doi.org/10.5281/zenodo.18107585;vi sit=swh:1:snp:d6b470221faf66d162688fca9dc9d3462d0f 425b;anchor=swh:1:rel:790599330ef5a22371a7f81df497 d4bd1037a86f;path=nayaksagar-PRELIMPFlow-752644f },
  version   = {v1.0.0},
}

@article{sakaguchi1993plugging,
  title={Plugging of the flow of granular materials during the discharge from a silo},
  author={Sakaguchi, Hide and Ozaki, Eiji and Igarashi, Tohru},
  journal={Int. J. Mod. Phys. B},
  volume={7},
  number={09n10},
  pages={1949--1963},
  year={1993},
  publisher={World Scientific}
}

\end{document}